\documentclass[twocolumn]{aastex631}

\usepackage[T1]{fontenc}

\usepackage{soul}

\usepackage[figuresright]{rotating}
\shorttitle{}
\shortauthors{Hasegawa et al.}
\graphicspath{{./}{figures/}}
\begin{document}

%\title{Mass Loss Rate as a Probe of Accretion and Dispersal Mechanisms for Protoplanetary Disks}
\title{Determining dispersal mechanisms of protoplanetary disks using accretion and wind mass loss rates}

%\correspondingauthor{Yasuhiro Hasegawa}
\email{yasuhiro.hasegawa@jpl.nasa.gov}

\author[0000-0002-9017-3663]{Yasuhiro Hasegawa}
\affiliation{Jet Propulsion Laboratory, California Institute of Technology, Pasadena, CA 91109, USA}

\author[0000-0002-9593-7618]{Thomas J. Haworth}
\affiliation{Astronomy Unit, School of Physics and Astronomy, Queen Mary University of London, London E1 4NS, UK}

\author[0000-0002-8636-3309]{Keri Hoadley}
\affiliation{Department of Physics \& Astronomy, University of Iowa, 203 Van Allen Hall, Iowa City, IA, 52242, USA}

\author[0000-0001-6072-9344]{Jinyoung Serena Kim}
\affiliation{Steward Observatory, Department of Astronomy, University of Arizona, 933 North Cherry Avenue, Tucson, AZ 85721, USA}
\affiliation{Alien Earths Team, NASA Nexus for Exoplanet System Science}

\author{Hina Goto}
\affiliation{Steward Observatory, Department of Astronomy, University of Arizona, 933 North Cherry Avenue, Tucson, AZ 85721, USA}

\author{Aine Juzikenaite }
\affiliation{Astronomy Unit, School of Physics and Astronomy, Queen Mary University of London, London E1 4NS, UK}

\author[0000-0001-8292-1943]{Neal J. Turner}
\affiliation{Jet Propulsion Laboratory, California Institute of Technology, Pasadena, CA 91109, USA}

\author[0000-0001-7962-1683]{Ilaria Pascucci}
\affiliation{Lunar and Planetary Laboratory, University of Arizona, Tucson, AZ 85721, USA}
\affiliation{Alien Earths Team, NASA Nexus for Exoplanet System Science}

\author[0000-0002-3131-7372]{Erika T. Hamden}
\affiliation{Steward Observatory, Department of Astronomy, University of Arizona, 933 North Cherry Avenue, Tucson, AZ 85721, USA}

%Just adding her details in case Aine gets involved

%\email{yasuhiro.hasegawa@jpl.nasa.gov}

\begin{abstract}
Understanding the origin of accretion and dispersal of protoplanetary disks is fundamental for investigating planet formation.
Recent numerical simulations show that launching winds are unavoidable 
when disks undergo magnetically driven accretion and/or are exposed to external UV radiation. 
Observations also hint that disk winds are common.
We explore how the resulting wind mass loss rate can be used as a probe of both disk accretion and dispersal.
As a proof-of-concept study, we focus on magnetocentrifugal winds, MRI (magnetorotational instability) turbulence, and external photoevapotaion.
By developing a simple, yet physically motivated disk model and coupling it with simulation results available in the literature,
we compute the mass loss rate as a function of external UV flux for each mechanism.
We find that different mechanisms lead to different levels of mass loss rate,
indicating that the origin of disk accretion and dispersal can be determined, by observing the wind mass loss rate resulting from each mechanism.
This determination provides important implications for planet formation.
This work thus shows that the ongoing and future observations of the wind mass loss rate for protoplanetary disks
are paramount to reliably constrain how protoplanetary disks evolve with time and how planet formation takes place in the disks.

\end{abstract}


\section{Introduction} \label{sec:intro}

Protoplanetary disks are widely accepted as the birth place of planetary systems
and are known as highly dynamical objects, but how do they evolve with time?
It is well established that once disks form and are isolated from surrounding environments,
disk evolution is caused by both mass accretion onto the host star and gas dispersal via launching winds from disk surfaces,
if complexity arising from planet formation is neglected \citep[e.g.,][]{2011ARA&A..49...67W,2016ARA&A..54..135H}.
One can therefore address this important question by identifying the dominant accretion and dispersal mechanisms in protoplanetary disks.

\begin{figure}%[!ht]
\begin{center}
\includegraphics[width=8.3cm]{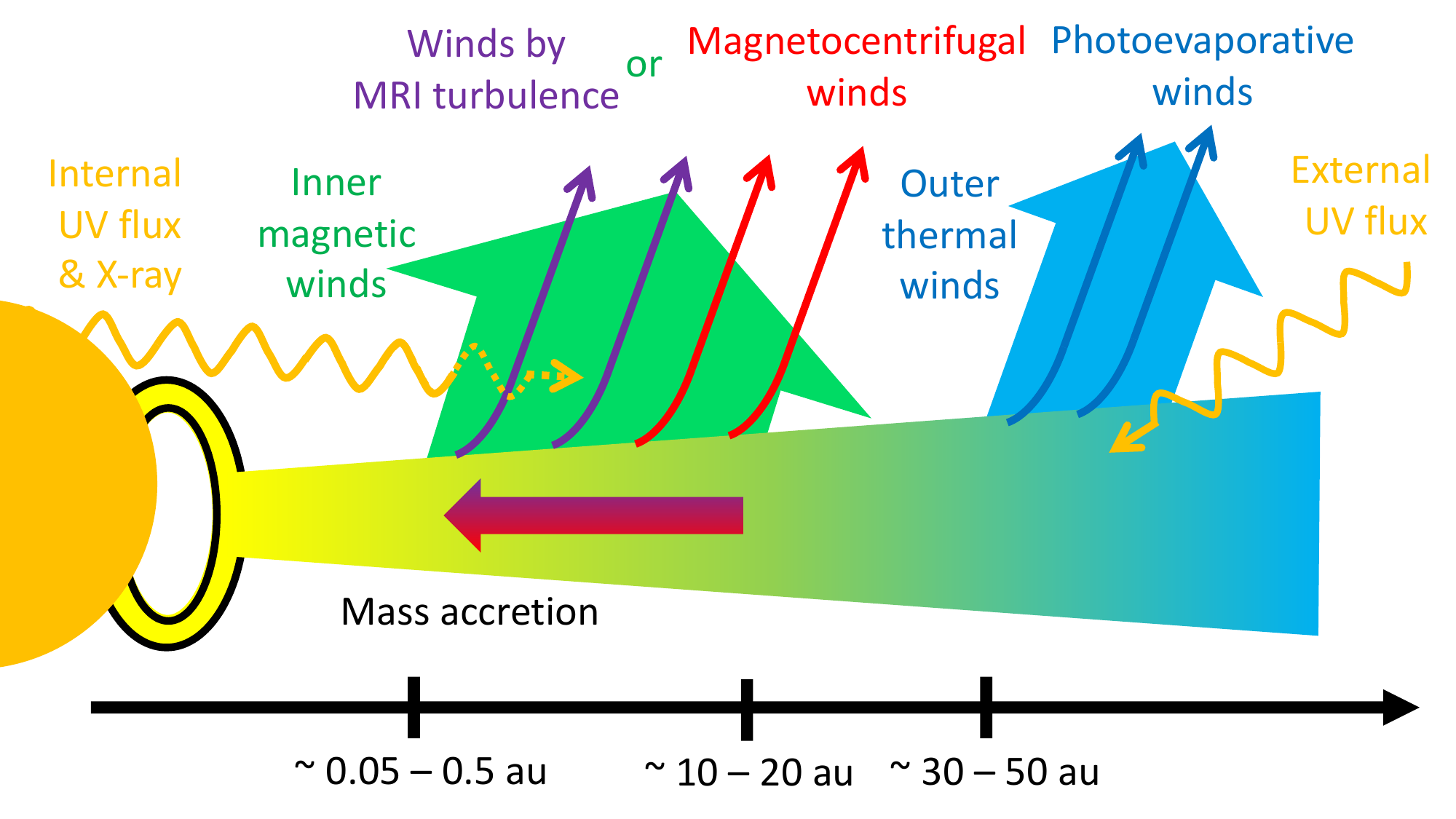}
\caption{Schematic diagram of a protoplanetary disk and the wind mechanisms examined in this work. 
We consider only three channels (mass accretion, inner magnetic winds, and outer thermal winds) for disk accretion and dispersal;
internal photoevaporation can be prevented due to the inner (massive) winds.
The inner magnetic winds should be radially extended to reproduce the observed accretion rate.
}
\label{fig1}
\end{center}
\end{figure}

A number of mechanisms have been proposed in the literature \citep[e.g.,][]{2014prpl.conf..411T,2017RSOS....470114E}.
Despite their variety, 
most mechanisms essentially assume that 
either magnetic fields threading through disks or heating by the host star and surrounding stars (or both) are responsible for disk accretion and dispersal.
For the former, coupling magnetic fields with the disk gas leads not only to angular momentum transport/removal, but also to launching winds 
\citep[e.g.,][]{2009ApJ...691L..49S,2013ApJ...769...76B}.
For the latter, heated gas gains the kinetic energy that fuel winds, 
especially in the outer disk, where the disk gas is less bound \citep[e.g.,][]{2004ApJ...611..360A,2019MNRAS.485.3895H}. 
Recent observations infer the presence of magnetically driven winds originating from inner ($\la 10$ au) disks 
\citep[e.g.,][]{2018ApJ...868...28F,2021ApJ...913...43W},
and these winds may be launched at even larger disk radii \citep[e.g.,][]{2020A&A...634L..12D}.
Thus, it may be plausible to consider that disk accretion and dispersal occur mainly through three channels (Figure \ref{fig1}):
mass accretion of disk gas onto its host star, inner magnetically-driven winds, and outer thermal winds.

In this Letter, we adopt the above theoretically motivated and observationally supported picture of disk evolution, 
examining how the mass loss rate resulting from the inner/outer winds can serve as a useful probe of both disk accretion and dispersal mechanisms. 
Given that the inner and outer winds can be distinguished observationally (e.g., by launching locations or spectral features),
we assume that the resulting mass loss rate is a local quantity computed from each wind.
As a proof-of-concept study, we consider only magnetocentrifugal winds (MCW), magnetorotational instability (MRI) turbulence, and external photoevapotaion (EPE; Figure \ref{fig1}), 
where the last one becomes important when disks are exposed to sufficiently strong external ultraviolet (UV) fields. 
In this work, we do not include internal photoevaporation (Section \ref{sec:discu}).
We show below that the wind mass loss rate arising from different mechanisms covers different parameter space.
Therefore, we will conclude that 
the wind mass loss rate measured by observations can reveal the origin of disk accretion and dispersal in a wide range of star-forming regions.

%\begin{figure*}
%\begin{minipage}{17cm}
%\end{minipage}
%\end{figure*}

\section{Disk accretion and dispersal} \label{sec:model}

The fundamental assumption of this work is that 
the ultimate origin of angular momentum transport/removal in protoplanetary disks are magnetic fields threading the disks,
and that at least disk surfaces are ionized enough to couple with the fields due to radiation from the host star and/or external radiation.

\subsection{Disk model}

We begin with the introduction of our disk model.
The conservation of mass and angular momentum allows one to obtain the general, mathematical expression of 
the mass accretion rate onto the host star \citep[$\dot{M}_{\rm acc}$, e.g.,][]{1998RvMP...70....1B,2016ApJ...821...80B,2016A&A...596A..74S}.
The expression assures decomposition of $\dot{M}_{\rm acc}$ into two components \citep[e.g.,][]{2017ApJ...845...31H}
\begin{equation}
\label{eq:mdot}
\dot{M}_{\rm acc}  = \dot{M}^{\rm Tur}_{\rm acc} + \dot{M}^{\rm DW}_{\rm acc},                    
\end{equation}
where $\dot{M}^{\rm Tur}_{\rm acc}$ and $\dot{M}^{\rm DW}_{\rm acc}$ denote the contributions arising from MHD turbulence and disk winds, respectively.
Many ideal and non-ideal MHD simulations confirm that the accretion stress is dominated by the Maxwell stress, rather than the Reynold stress
\citep[e.g.,][]{1995ApJ...440..742H,2013ApJ...769...76B}.
Therefore, $\dot{M}^{\rm Tur}_{\rm acc}$ and $\dot{M}^{\rm DW}_{\rm acc}$ are specified once the strength and geometry of magnetic fields are determined.

When the $\alpha-$prescription is used \citep{1973A&A....24..337S},
the effective viscosity ($\alpha_{\rm SS}$) mimicking disk turbulence is given as \citep[e.g.,][]{2017ApJ...845...31H}
\begin{equation}
\label{eq:alpha_eff}
\alpha_{\rm SS} %= \frac{1}{3 \pi} \frac{H_{\rm w}  \langle \overline{B_r B_{\phi}}\rangle}{ \Sigma_{\rm g} c_{\rm s}^2}
                         =  \frac{1}{3 \pi} \frac{ \dot{M}^{\rm Tur}_{\rm acc} }{ \Sigma_{\rm g} H_{\rm g}^2 \Omega},
\end{equation}
where $\Omega=\sqrt{GM_{\rm s}/r^3}$ is the Keplerian, angular frequency around the host star with the mass of $M_{\rm s}$,
$\Sigma_{\rm g}$ the gas surface density, $H_{\rm g}= c_{\rm s}/ \Omega$ the gas pressure scale height, 
and $c_{\rm s}$ the sound speed of the disk gas.

In the following, the above disk model is used to compute the mass loss rate
with $M_{\rm s} = 1M_{\odot}$, $H_{\rm g} / r = h_0(r/ 1\mbox{ au})^{1/4}$ , and $h_0=0.05$.

\subsection{Magnetocentrifugal winds} \label{sec:mcw}

The recent progress in non-ideal MHD simulations suggests the importance of magnetocentrifugal winds in protoplanetary disks \citep[e.g.,][]{2014prpl.conf..411T};
given that protoplanetary disks are generally dense and cold,
MRI and the resulting MHD turbulence are likely quenched in most regions 
\citep[e.g., ][]{1996ApJ...457..355G,2007Ap&SS.311...35W}.
Magnetocentrifugal winds offer an alternative mechanism of disk accretion.

The primary origin of these winds is magnetocentrifugal force.
The force becomes effective when disks are threaded by relatively strong, open  magnetic fields that are suitably inclined from the polar axis;
the differential rotation of Keplerian disks winds up the poloidal component of magnetic fields above the disk surfaces.
These fields act like a lever arm anchored on the disk surfaces and can efficiently remove disks' angular momentum by launching winds.

The corresponding mass loss rate ($\dot{M}_{\rm loss}^{\rm MCW}$) is computed from the conservation of angular momentum \cite[e.g.,][]{2016ApJ...818..152B}:
\begin{equation}
\label{eq:mag_cen}
\left. \dot{M}_{\rm acc}^{\rm DW} \frac{dj}{dr} \right|_{r_{\rm w}} = \frac{d \dot{M}_{\rm loss}^{\rm MCW}}{dr} \Omega (r_{\rm A}^2-r_{\rm w}^2),
\end{equation}
where $j(r) \equiv \Omega r^2$ is disks' specific angular momentum at $r$, 
$r_{\rm w}$ the wind launching radius, and $r_A$ the Alfv\'{e}n radius.
Note that $\dot{M}_{\rm loss}^{\rm MCW}$ is the {\it cumulative} mass loss rate from the wind launching region, that is
\begin{equation}
\dot{M}_{\rm loss}^{\rm MCW} = \int_{r_{\rm in}}^{r_{\rm out}} dr 2 \pi r (\rho v)_{H_{\rm w}}^{\rm MCW},
\end{equation}
where $r_{\rm in}$ and $r_{\rm out}$ are the inner and outer boundaries of the region, and 
$(\rho v)_{H_{\rm w}}^{\rm MCW}$ are the gas density and velocity at the wind base wth a height of $H_{\rm w}$, respectively. 

Equation (\ref{eq:mag_cen}) is rewritten as
\begin{equation}
\label{eq:xi}
\xi (r_{\rm A}/r_{\rm w}) \equiv \left. \frac{d \dot{M}_{\rm loss}^{\rm MCW} / d \ln r}{\dot{M}_{\rm acc}^{\rm DW}} \right|_{r_{\rm w}} = \frac{1}{2} \frac{1}{(r_{\rm A}/r_{\rm w})^2 -1},
\end{equation}
which is referred to as the "ejection index" \citep{1995A&A...295..807F}.
Also, $(r_{\rm A}/r_{\rm w})$ is often called as the "magnetic lever arm".
This equation indicates that $r_{\rm A}/r_{\rm w} \geq \sqrt{3/2} \simeq 1.2$ since $\xi (r_{\rm A}/r_{\rm w}) \leq1$.\footnote{
Recent non-ideal MHD simulations show that  $r_{\rm A}/r_{\rm w}$ can become smaller than 1.2 \citep{2017ApJ...845...75B,2019ApJ...874...90W};
in this case, the primary origin of launching winds is likely the magnetic pressure arising from the toroidal component.
Given that such winds tend to be massive due to inefficient angular momentum removal,
magnetocentrifugal winds become more important for comparing with the MRI turbulence case  (see Section \ref{sec:comp}).}
Based on the results of non-ideal MHD simulations \citep{2020ApJ...896..126G},
we adopt that $r_{\rm A}/r_{\rm w} \simeq 1.6$ as an upper limit (see their table 2).

Consequently, the mass loss rate due to magnetocentrifugal winds is given as
\begin{equation}
\label{eq:dotM_loss_mcf}
\dot{M}_{\rm loss}^{\rm MCW}  = \dot{M}_{\rm acc}^{\rm DW}  \int_{r_{\rm in}}^{r_{\rm out}} dr \frac{\xi (r_{\rm A}/r) }{r}.
\end{equation}

\subsection{MRI turbulence} \label{sec:mri}

MRI has been well recognized as the central engine of disk accretion
because the instability surely grows in Keplerian disks \citep{1991ApJ...376..214B}.
As described above, however, its operation in protoplanetary disks has recently been challenged due to non-ideal MHD effects,
except for the vicinity of the host star.
It is thus of fundamental importance to determine whether MRI-driven turbulence plays a dominant role in disk evolution.

The wind mass loss rate can be used as a tracer of MRI turbulence because ideal MHD simulations show that MRI turbulence can launch winds
(\citealt{2009ApJ...691L..49S}, c.f., see \citealt{2000ApJ...534..398M}).
The origin of winds is the generation of large scale channel flow in the vertical direction and its subsequent breakup due to magnetic reconnection around the disk surface.
The reconnection converts magnetic energy to thermal energy \citep[e.g.,][]{2001ApJ...561L.179S},
and hence the ultimate source of energy to launch winds is the gravitational energy released by the accretion stress, 
phenomenologically the viscous heating \citep{2016A&A...596A..74S}.\footnote{
The operation of this mechanism is confirmed even when ohmic resistivity is included \citep{2010ApJ...718.1289S};
the presence of MRI-active surface layers leads to the production and breakup of large scale channel flow around disk surfaces.}

The energy available for winds can be constrained from the conservation law \citep{2016A&A...596A..74S}.
The resulting constraint leads to the condition that $0\leq 1/\xi \leq 1$; equivalently, $1 \leq r_{\rm A}/r_{\rm w} \leq \sqrt{3/2} \simeq 1.2$ 
(see Appendix \ref{sec:app1} for mathematical derivation).
This condition ensures that MRI turbulence and magnetocentrifugal winds are mutually exclusive.

A tighter constraint is obtained by MHD simulations;
\citet{2010ApJ...718.1289S} show that $(\rho v)_{H_{\rm w}}^{\rm MRI} / (\rho_{\rm mid} c_{\rm s})$ is linearly proportional to $\alpha_{\rm SS}$ broadly,
where $\rho_{\rm mid} = \Sigma_{\rm g} / (\sqrt{2 \pi}H_{\rm g})$ is the gas density at the midplane (see their figure 2).
This linear relation is the direct reflection that winds are launched due to the accretion energy
and allows one to explicitly define the energy loss ($E^{\rm MRI}_{\rm w} $) caused by winds due to MRI turbulence as (see Appendix \ref{sec:app1})
\begin{equation}
\label{eq:E_py2}
E^{\rm MRI}_{\rm w} \equiv \frac{r^2 \Omega^2}{2}  (\rho v)_{H_{\rm w}}^{\rm MRI} 
                                        = \frac{\epsilon}{(2\pi)^{3/2}} \frac{\dot{M}_{\rm acc}^{\rm Tur}}{H_{\rm g}^2} \frac{r^2 \Omega^2}{2}, 
\end{equation}
where $\epsilon$ is the proportionality constant determined by MHD simulations.
Then, the energy partition coefficient ($\eta$) of the accretion energy between winds and radiation is given as
\begin{equation}
\label{eq:eta}
\eta (\epsilon, r) \equiv \frac{ E^{\rm MRI}_{\rm w} }{3 \Omega^2 \dot{M}_{\rm acc}^{\rm Tur} / (4 \pi)}= \frac{ \epsilon}{3 \sqrt{2 \pi}} \left( \frac{r}{H_{\rm g}} \right)^2.
\end{equation}
Based on the simulation results of \citet{2009ApJ...691L..49S}, 
$\alpha =0.012$ and $(\rho v)_{H_{\rm w}}^{\rm MRI} / (\rho_{\rm mid} c_{\rm s}) \simeq 8 \times 10^{-5}$ at the plasma $\beta=10^6$ with $\sqrt{2}H_{\rm g} /r =0.1$.
This leads to that $\eta (\epsilon \simeq 7 \times 10^{-3},r) \simeq 0.18$,
suggesting that about 18 \% of the accretion energy is used to launch winds.
Note that this estimate should be an upper limit;
the simulations adopt the shearing box approximation,
and the isothermal assumption is employed,
both of which increases $(\rho v)_{H_{\rm w}}^{\rm MRI}$.
We therefore consider that the following may be a reasonable upper limit:
$\eta = 0.1$; equivalently, $\epsilon \simeq 2 \times 10^{-3}$ at $r=1$ au.

As a result, the wind mass loss rate due to MRI turbulence is written as (see equations (\ref{eq:E_py2}) and (\ref{eq:eta})) 
\begin{eqnarray}
\label{eq:dotM_loss_mri}
\dot{M}_{\rm loss}^{\rm MRI} & = &  \int_{r_{\rm in}}^{r_{\rm out}} dr 2 \pi r (\rho v)_{H_{\rm w}}^{\rm MRI}  \\ \nonumber
                                              & = &  \int_{r_{\rm in}}^{r_{\rm out}} dr \dot{M}_{\rm acc}^{\rm Tur} \frac{3 \eta (\epsilon, r)}{r} % \\ \nonumber
                                              % & = & 
                                               = \frac{ \dot{M}_{\rm acc}^{\rm Tur} }{ \sqrt{2 \pi} } \int_{r_{\rm in}}^{r_{\rm out}} dr  \frac{\epsilon}{r} \left( \frac{r}{H_{\rm g}} \right)^2.
                                              %& \simeq &  \frac{  H_{\rm w}}{ \sqrt{2 \pi} \Omega}  \langle \overline{B_r B_{\phi}}\rangle  \int_{r_{\rm in}}^{r_{\rm out}} dr \frac{ \epsilon}{r} \left( \frac{r}{H_{\rm g}} \right)^2.
\end{eqnarray}

\subsection{External photoevaporation} \label{sec:epe}

Massive stellar clusters host OB stars that emit strong UV radiation. 
If the UV field incident upon a disk is sufficiently high, it can drive winds from the outer disk \citep[e.g.,][]{2000ApJ...539..258R,2019MNRAS.485.3895H}.
The process is known as external photoevaporation and contributes to disk dispersal significantly.

We compute external photoevaporative mass loss rates ($\dot{M}_{\rm loss}^{\rm EPE}$), by utilizing the \textsc{fried} grid \citep{2018MNRAS.481..452H}; 
\textsc{fried} is built upon from the results of radiation hydrodynamic models and 
provides the mass loss rate as a function of the stellar mass ($M_{\rm s}$), the disk mass ($M_{\rm d}$) and radius ($r_{\rm d}$), and external far UV (FUV) flux ($F_{\rm FUV}$).\footnote{
This work adopts plausible ranges of $M_{\rm d}$ and $r_{\rm d}$ used in \textsc{fried} 
(i.e., $10^{-3} \le M_{\rm d}/M_{\odot} \le 10^{-1}$ and $30 \mbox{ au} \le r_{\rm d} \le 400 \mbox{ au}$, see Appendix \ref{sec:app5} for a parameter study.)
Also, $F_{\rm FUV}$ is in units of $G_0$, 
where 1 $G_{0} =1.6 \times 10^{-3}$ erg cm$^{-2}$ s$^{-1}$ is the flux integrated over the range from 6 to 13.6 eV \citep{1968BAN....19..421H}.
The background $F_{\rm FUV}$ is about $1~G_0$ in the solar neighborhood.}
Previous studies show that $\dot{M}_{\rm loss}^{\rm EPE}$ is a strong function of $r_{\rm d}$, 
more easily stripping material from the outer (less bound) regions of large disks \citep[e.g.,][]{2004ApJ...611..360A}. 
When the mass loss rate is higher than the rate of viscous spreading, 
the disk is truncated \citep[e.g.][]{2007MNRAS.376.1350C}. 
This occurs rapidly until the disk is shrunk to a radius at which $\dot{M}_{\rm loss}^{\rm EPE}$ becomes comparable to $\dot{M}_{\rm acc}$ 
\citep[also see Appendix \ref{sec:app2} for mathematical confirmation]{2020MNRAS.497L..40W}.
Thus, $\dot{M}_{\rm loss}^{\rm EPE} \geq \dot{M}_{\rm acc}$ for disks undergoing accretion. 
Note that external photoevaporation does not transport/remove disks' angular momentum,
and hence accretion should be driven by magnetic fields as in Sections \ref{sec:mcw} and \ref{sec:mri}.
This suggests that disks may exhibit both the inner magnetic and outer thermal winds simultaneously.

We compile the \textsc{fried} grid to find out the maximum value of $\dot{M}_{\rm loss}^{\rm EPE}$ as functions of $r_{\rm d}$ and $F_{\rm FUV}$.
As discussed below, our focus is on the ratio of the wind mass loss rate to the stellar accretion rate.
We therefore convert $M_{\rm d}$ to $\dot{M}_{\rm acc}$, using the observed correlation between $M_{\rm d}$ and $\dot{M}_{\rm acc}$ \citep{2019A&A...631L...2M}:
\begin{equation}
\label{eq:corr_Macc_Md}
\log \left( \frac{ \dot{M}_{\rm acc} }{M_{\odot} \mbox{ yr}^{-1} } \right) = \left( 0.9 \pm 0.1 \right) \log \left( \frac{M_{\rm d}}{M_{\odot}} \right) -( 6.5 \pm 0.4).
\end{equation}
Note that this correlation is derived originally from the dust disk mass (not the gas disk mass).
Conversion from the dust to gas mass is done with the assumption that the gas-to-dust ratio is 100.
Also, the observed data are obtained from objects in Chameleon I and Lupus, which both reside in low UV environments;
the estimated value of $\dot{M}_{\rm acc}$ becomes smaller for objects in high UV environments,
given that disks in such environments are fully or partially truncated by external photoevaporation and hence are less massive.
This trend is indeed confirmed by \citet[see their figure 7]{2017MNRAS.468.1631R}.
We however use the correlation because the counterpart in high UV environments is not available in the literature currently;
our estimate should be viewed as a conservative one.

%\begin{figure*}
%\begin{minipage}{17cm}
\begin{figure}%[!ht]
\begin{center}
\includegraphics[width=8.3cm]{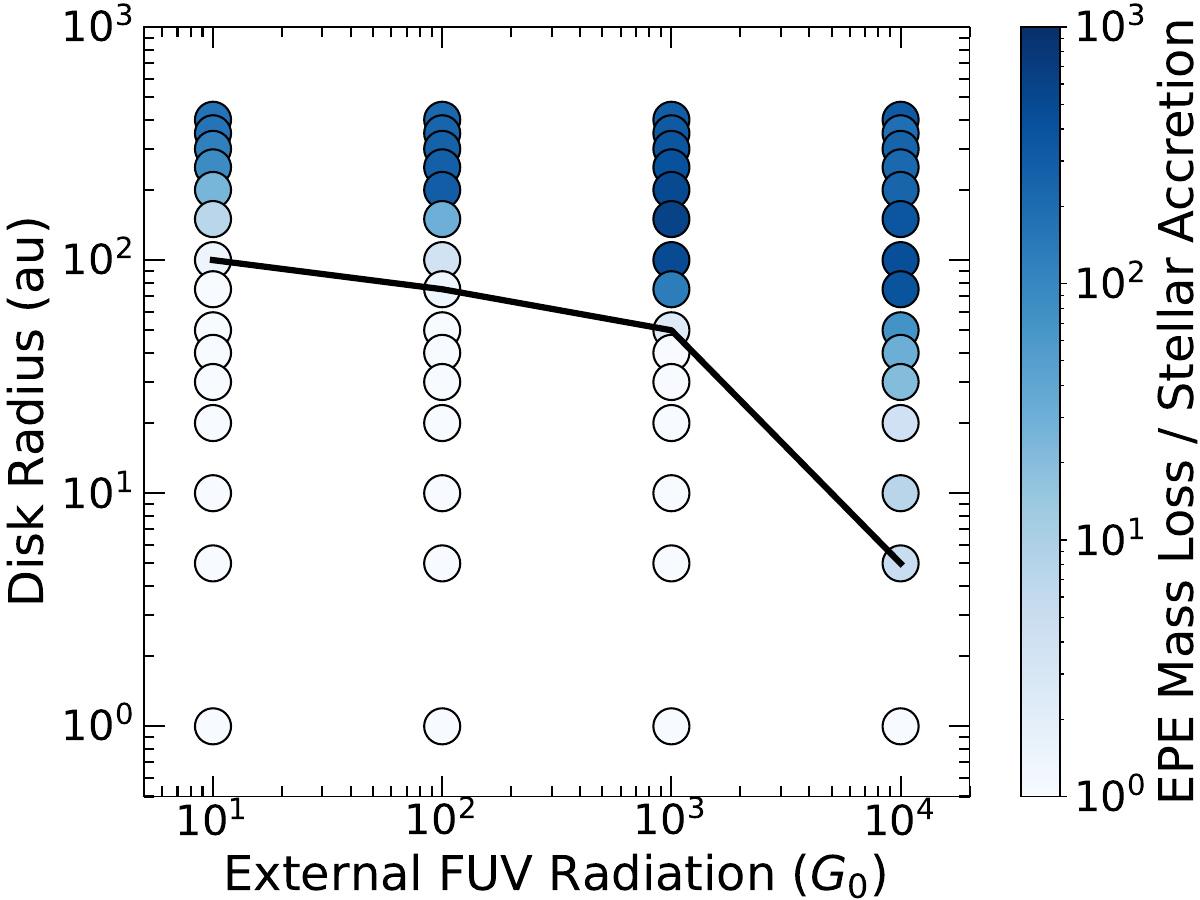}
\caption{The maximum value of the ratio between external photoevaporative mass loss rates and stellar accretion rates.
The \textsc{fried} grid is used to estimate the mass loss rate, 
and the observed correlation is used to convert the disk mass to the stellar accretion rate.
The ratio is an increasing function of disk radius and external FUV radiation.
The black solid line defines the region where the ratio becomes larger than unity;
below the line, disks are considered to have been already truncated by external photoevaporation.}
\label{fig2}
\end{center}
\end{figure}
%\end{minipage}
%\end{figure*}

Figure \ref{fig2} shows the resulting, maximum value of $\dot{M}_{\rm loss}^{\rm EPE}/\dot{M}_{\rm acc}$.
As expected, the ratio becomes higher for larger $r_{\rm d}$, 
and disk truncation (that is, $\dot{M}_{\rm loss}^{\rm EPE}/\dot{M}_{\rm acc} \simeq 1$) occurs for smaller-sized disks with stronger $F_{\rm FUV}$.

In the following, we use Figure \ref{fig2} to compare $\dot{M}_{\rm loss}^{\rm EPE}/\dot{M}_{\rm acc}$ with other dispersal mechanisms.

\subsection{Comparison} \label{sec:comp}

\begin{table*}
\begin{minipage}{17cm}
%\begin{table}
\begin{center}
\caption{The ratio of the wind mass loss rate to the stellar accretion rate for each mechanism}
\label{table1}
{\renewcommand{\arraystretch}{1.2}
%{\scriptsize
\begin{tabular}{l||c|c}
\hline
Origin                                                &  Parameter                                                                                  & Ratio                                                                                                                        \\   \hline \hline         
Magnetocentrifugal winds (MCW)    &  $1.2 \la r_{\rm A}/r  \la 1.6$ ($0.32 \la \xi  \la 1$)                        & $0.74 \la \dot{M}_{\rm loss}^{\rm MCW}/\dot{M}_{\rm acc}^{\rm MCW} \la 2.3 $     \\      %\hline 
MRI turbulence (MRI)                      &  $ \epsilon \leq 2 \times 10^{-3}$ ($ \eta \leq  0.1$ at $r=1$ au)  & $\dot{M}_{\rm loss}^{\rm MRI} /\dot{M}_{\rm acc}^{\rm MRI} \la 0.44$                      \\                                                                   
External Photoevaporation (EPE)   &   Figure \ref{fig2}                                                                           &  $1 \la \dot{M}_{\rm loss}^{\rm EPE} / \dot{M}_{\rm acc} \la 600$                \\                                                                 
 \hline 
\end{tabular}
%}
}
\end{center}
%\end{table}
\end{minipage}
\end{table*}

We now determine whether the wind mass loss rate can be used as a probe of the accretion and dispersal mechanisms for protoplanetary disks.
To directly compare the various dispersal mechanisms discussed above, we compute the ratio ($\dot{M}_{\rm loss}/\dot{M}_{\rm acc}$),
under the assumption that in each case, only the corresponding mechanism plays the dominant role for both $\dot{M}_{\rm acc}$ and $\dot{M}_{\rm loss}$.
We constrain the parameter space in which each mechanism becomes most important, 
by specifying the upper and lower limits of $\dot{M}_{\rm loss}/\dot{M}_{\rm acc}$.

Table \ref{table1} summarizes the results.
For the MCW and MRI cases,
we set that $r_{\rm in}= 1$ au and $r_{\rm out} = 10$ au,
following recent observations \citep[e.g.,][]{2018ApJ...868...28F,2018A&A...618A.120L,2021ApJ...913...43W}.
These choices are intended to examine how two kinds of inner, magnetically driven winds are differentiated by observations,
and do not necessarily mean that the winds should be launched only in the region.
For the EPE case, the widest range is picked in Table \ref{table1} for reference purpose,
which is achieved at $10^3 G_0$ (see Figure \ref{fig2}).

%\begin{figure*}
%\begin{minipage}{17cm}
\begin{figure}%[!ht]
\begin{center}
\includegraphics[width=8.3cm]{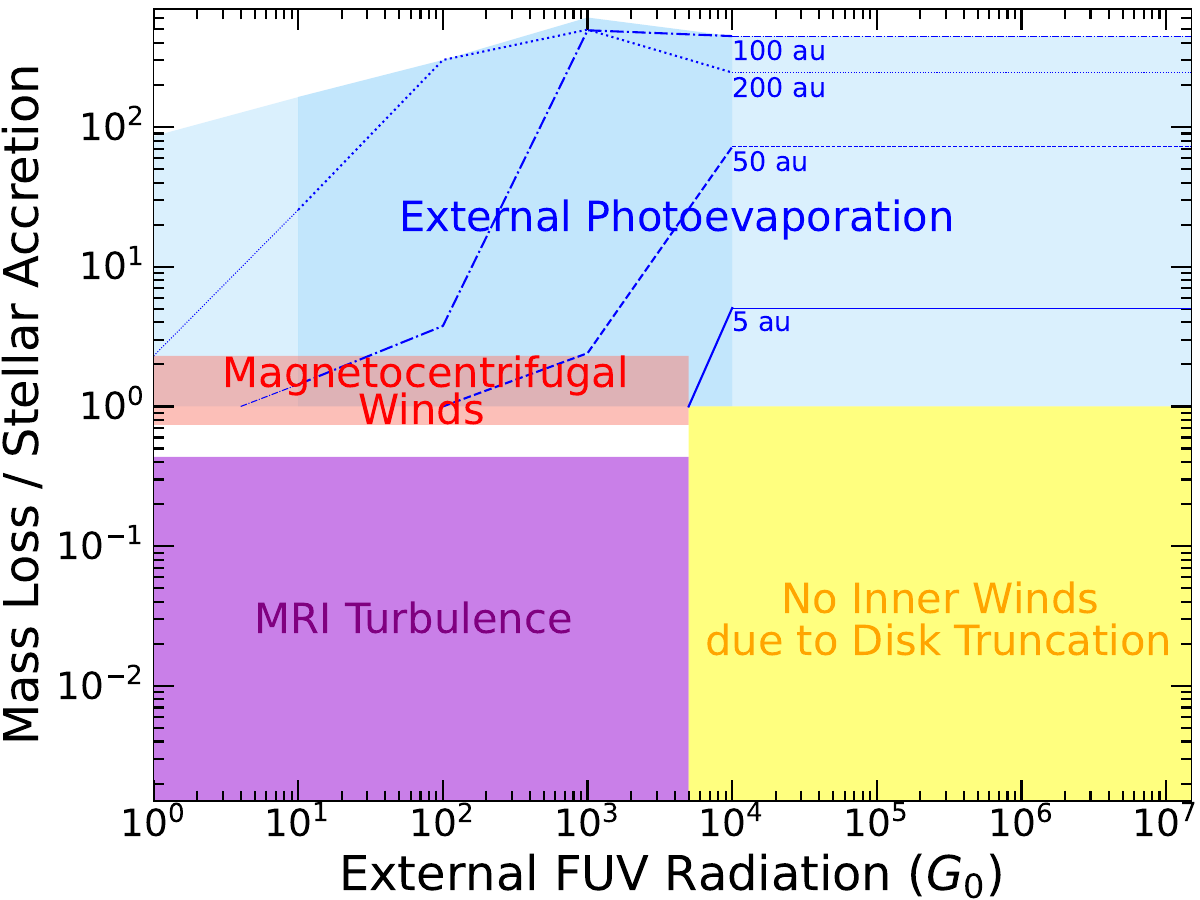}
\caption{The ratio of the wind mass loss rate to the stellar accretion rate as a function of the external UV flux (also see Table \ref{table1} and Figure \ref{fig2}).
The MCW case is denoted by the red shaded region, the MRI case the purple shaded region, and the EPE case the blue shaded region.
For the EPE case, the disk radius dependence is explicitly denoted by the lines.
The parameter space that is not covered by the current \textsc{fried} grid is denoted by the light blue shaded regions.
Two inner magnetic winds can be differentiated by measuring the resulting mass loss rate,
and the outer thermal winds and the resulting mass loss rate can be constrained by observing the disk radius.}
\label{fig3}
\end{center}
\end{figure}
%\end{minipage}
%\end{figure*}

Figure \ref{fig3} shows the synthesized results.
We find that the inner magnetic winds (i.e., MCW and MRI) play the major role in disk's angular momentum transport/removal at low UV environments,
and their resulting mass loss can be observed at $F_{\rm FUV} \la 5 \times 10^3 G_0$; 
beyond this value, external photoevaporation truncates disks so much that the wind launching region disappears (see the yellow shaded region).
It is important that the MCW and MRI cases occupy different parameter space.
Our results therefore indicate that if the corresponding mass loss and accretion rates are measured accurately enough,
the origin of angular momentum transport/removal is determined.
Note that time variability of mass loss rates may also be used to differentiate two kinds of inner magnetic winds \citep[e.g.,][]{2010ApJ...718.1289S}.

For the EPE case, the current \textsc{fried} grid covers only from $10 G_0$ to $10^4G_0$,
and we extrapolate the results toward $1 G_0$ and put a limit for $> 10^4G_0$ (see the light blue shaded regions);
both are reasonable because at lower UV environments, the effect of photoevaporation becomes weaker,
while at very high UV environments, disk truncation occurs very rapidly.
The latter is indeed confirmed in Figure \ref{fig3};
the value of $\dot{M}_{\rm loss}^{\rm EPE}$ with $r_{\rm d} = 100$ au becomes higher than that with $r_{\rm d} = 200$ au at $F_{\rm FUV} > 10^3 G_0$
because truncation is already in process for larger disks.
We find that the resulting mass loss rate becomes very sensitive to the disk radius, as expected.

In summary, the mass loss rate resulting from inner magnetic winds can be used to determine the origin of angular momentum transport/removal 
and hence the dynamical properties (turbulent vs laminar) of the disk gas.
This is the fundamental information for constraining how planets form in the disks (see Section \ref{sec:discu}).
The mass loss rate originating from outer thermal winds is a sensitive function of the disk radius.
Most of the disk mass distribute in the outer part of the disk, 
and hence external photoevaporation regulates the formation timescale of planets and their final mass.

\section{Discussion} \label{sec:discu}

This work has so far considered a specific set of parameters.
We have conducted a parameter study and confirmed that the variation of parameters does not affect our findings significantly (see Appendix \ref{sec:app5});
the MRI-dominated region can extend and overlap with the MCW one, (that is, the empty space in Figure \ref{fig3} may disappear).
However, the parameter space specified in Figure \ref{fig3} well represents the population of certain types of winds, 
and hence differentiating the origin of disk winds in the parameter space is possible, by observing a good number of disks.
Our model has also assumed that disk surfaces are ionized enough to couple with magnetic fields,
which is requisite both for MCWs and MRI.
The ionization structure of disks is poorly understood.
The presence of (massive) inner winds may prevent the host star's photons from reaching the disk surface and ionizing it \citep{2020ApJ...903...78P},
while external heating (e.g., cosmic ray) may be important for disk ionization \citep{2021ApJ...912..136S}.
Recent observations suggest that MRI is unlikely to operate at the surface of some disks \citep{2020ApJ...895..109F},
which may be used to constrain disk ionization.
Finally, we have assumed that photoevaporation occurs only via external UV radiation.
This is based on recent observations which propose that the presence of inner (massive) winds can shield high energy radiation from the host stars
and suppress internal photoevaporation \citep[see Figure \ref{fig1}]{2020ApJ...903...78P}.
More observations are obviously needed to verify (or falsify) this picture;
if internal photoevaporation drives winds, the corresponding mass loss rate should be comparable to that of MRI.
Identifying the wind launching region may allow one to differentiate these two winds.

\begin{figure*}
\begin{minipage}{17cm}
%\begin{figure}%[!ht]
\begin{center}
\includegraphics[width=15cm]{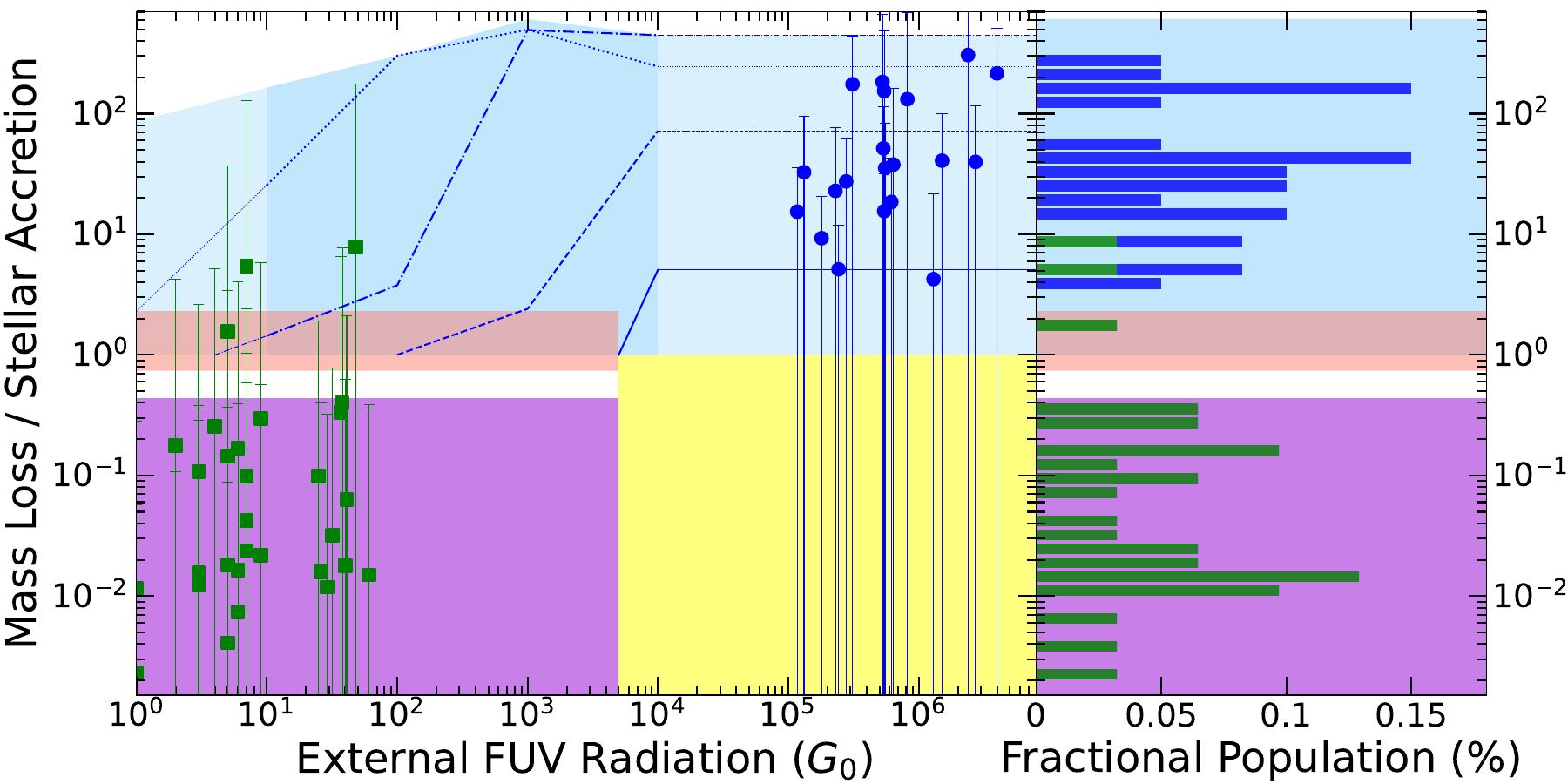}
\caption{The distribution of the data currently available in the literature and the fractional population on the left and right panels, respectively.
The observed systems are summarized in Tables \ref{table_app3} and \ref{table_app4} (denoted by the green and blue points, respectively), 
and the error estimation is described in Appendices \ref{sec:app3} and \ref{sec:app4}.
The current, large uncertainties prevent one from determining the dominant accretion and dispersal mechanisms,
while the data already show some structure, distributing mainly in the MRI and EPE regions.}
\label{fig4}
\end{center}
%\end{figure}
\end{minipage}
\end{figure*}

Measuring the wind mass loss rate of protoplanetary disks provides a number of important implications for planet formation.
The current observations are not accurate enough due to large uncertainties (Figure \ref{fig4}).
However, the ongoing and future observations will improve accuracy.
The most impactful implication of such observations is that they can constrain the mechanism of angular momentum transport/removal in protoplanetary disks.
It is one fundamental parameter needed in planet formation models at all the stages: 
dust growth is limited by disk turbulence, especially in the inner disk region \citep[e.g.,][]{2012A&A...539A.148B};
planetesimal dynamics and growth are very sensitive to disk turbulence \citep[e.g.,][]{2008ApJ...686.1292I}; 
the formation of planetary cores by pebble accretion is a function of the disk turbulence \citep[e.g.,][]{2016ApJ...825...63C}; 
and the behavior of planetary migration can be drastically altered by the level of disk turbulence \citep[e.g.,][]{2019MNRAS.484..728M}.
Cumulatively, the resulting population of planets may differ significantly \citep[e.g.,][]{2021arXiv211101798S}.
Note that when non-ideal MHD effects quench MHD turbulence, 
hydrodynamical instabilities come into play to excite hydrodynamical turbulence in disks \citep[][]{2019PASP..131g2001L}.

Another important implication is that measuring the mass loss rate in different UV environments allows one to explore 
when external photoevaporation plays a defining role in disk evolution and planet formation. 
This will provide new insights about how sensitive planet formation is to the surrounding environments, 
and shed light on the origin of the diversity of observed exoplanetary systems \citep[e.g.,][]{2019MNRAS.490.5678C, 2020Natur.586..528W}. 

In conclusion, observing wind mass loss rates of protoplanetary disks in a wide range of star-forming regions provides a key diagnostic of disk accretion and dispersal mechanisms.
Conducting such observations is one critical step to better understand how protoplanetary disks evolve with time and how planet formation takes place in the disks.

\begin{acknowledgments}

The authors thank an anonymous referee for the useful comments on the manuscript.
The research was carried out at the Jet Propulsion Laboratory, California Institute of Technology, under a contract with the National Aeronautics and Space Administration (80NM0018D0004).
Y.H. is supported by JPL/Caltech.       

\end{acknowledgments}

\bibliographystyle{aasjournal}
\bibliography{adsbibliography}    %% includes the journal abbrevs

\appendix

\section{The energy budget for winds due to MRI turbulence} \label{sec:app1}

The conservation of energy constrains the energy available for winds due to MRI turbulence \citep{2016A&A...596A..74S}:
\begin{equation}
\label{eq:ene_winds}
  \left.  \left( E_{\rm w} + \frac{r^2 \Omega^2}{2} \right) (\rho v_z)_{H_{\rm w}} +  F_{\rm rad} \right|^{H_{\rm w}}_{-H_{\rm w}}  
=  \frac{ \Omega^2}{4 \pi} \left( 3 \dot{M}_{\rm acc}^{\rm Tur} + \dot{M}_{\rm acc}^{\rm DW} \right),
\end{equation}
where $E_{\rm w}$ is the energy needed to launch winds, and $F_{\rm rad}$ is the energy loss due to radiation.
According to \citet{2016A&A...596A..74S}, we assume that $E_{\rm w}=0$, which is the minimum requirement.
In addition, the following amount of energy is carried away by magnetocentrifugal winds  (see equation (\ref{eq:xi})):
\begin{equation}
\frac{ \Omega^2}{4 \pi}  \dot{M}_{\rm acc}^{\rm DW} =  \frac{1}{\xi (r_{\rm A}/r) }  \frac{r^2 \Omega^2}{2}   (\rho v_z)_{H_{\rm w}}^{\rm MCW}.
\end{equation}
In other words, the energy loss ($E^{\rm MRI}_{\rm w}$) caused by winds due to MRI turbulence should have the range of, in the limit that $F_{\rm rad} =0$,
\begin{equation}
\label{eq:E_py}
 \left[ 1  - \frac{1}{\xi (r_{\rm A}/r) } \right] \frac{r^2 \Omega^2}{2} (\rho v_z)_{H_{\rm w}}  \leq E^{\rm MRI}_{\rm w} \leq  \frac{r^2 \Omega^2}{2} (\rho v_z)_{H_{\rm w}}.
\end{equation}
The above equation indicates that the necessary condition is $0\leq 1/\xi \leq 1$; equivalently, $1 \leq r_{\rm A}/r_{\rm w} \leq \sqrt{3/2} \simeq 1.2$.

Thus, it is guranteed that MRI turbulence and magnetocentrifugal winds are mutually exclusive in terms of the magnetic lever arm.
Note that the above range of $E^{\rm MRI}_{\rm w}$ should be viewed as an upper bound due to the limit that $F_{\rm rad} =0$.

\section{The mass loss rate for disks truncated by external photoevaporation} \label{sec:app2}

Efficient external photoevaporation truncates protoplanetary disks when disks are exposed to high external UV radiation long enough.
For this case, the wind mass loss rate is controlled by viscous spreading if disk accretion is driven by viscosity;
even when disks are already truncated,
viscous spreading expands their sizes, and some gas is moved to the region where external photoevaporation is effective.
\citet{2020MNRAS.497L..40W} already explore this phenomenon in detail.
We here provide a brief mathematical confirmation.

The mass loss rate ($\dot{M}_{\rm loss, tru}^{\rm EPE}$) for disks truncated by external photoevaporation is given as
\begin{equation}
\dot{M}_{\rm loss, tru}^{\rm EPE} = \zeta (r_{\rm t}) \dot{M}_{\rm acc}^{\rm Tur}, % \simeq \zeta (r_{\rm t}) \frac{  H_{\rm w}}{ \Omega}  \langle \overline{B_r B_{\phi}}\rangle,
\end{equation}
where $r_{\rm t}$ is the outer edge of truncated disks,
\begin{equation}
\zeta (r _{\rm t}) \equiv \frac{2 (2 + p)}{t_{\rm age}/t_{\rm vis } (r_{\rm t})+1} - 1,
\end{equation}
$p$ is the power-low index of $\Sigma_{\rm g} ( \propto r^p$), $t_{\rm age}$ is the age of the system,
and $t_{\rm vis}(r_{\rm t})$ is the local viscous timescale at $r=r_{\rm t}$.
Note that $r_{\rm t}$ is automatically determined once the value of $F_{\rm FUV}$ is specified.
We have used the similarity solution to obtain the functional form of $\zeta (r _{\rm t})$ \citep[e.g.,][]{1998ApJ...495..385H},
which is derived from the conservation of angular momentum
under the assumption that disks' angular momentum is transported by effective viscosity ($\nu$).

Given that the lifetime of truncated disks becomes comparable to $t_{\rm vis } (r_{\rm t})$
and the \textsc{fried} grid adopts that $p=-1$ \citep{2018MNRAS.481..452H},
$\zeta (r _{\rm t})$ can be re-written as
\begin{equation}
\zeta (r _{\rm t}) \simeq  \left\{ 
                           \begin{array} {l}
                                                 1         \ \       \mbox{ for } t_{\rm age} \la t_{\rm vis } (r_{\rm t}), \\
                                                 0         \ \       \mbox{ for } t_{\rm age} \ga  t_{\rm vis } (r_{\rm t}),
                           \end{array} 
                     \right.
\end{equation}
where $t_{\rm vis } (r_{\rm t})$ is given as
\begin{equation}
t_{\rm vis } (r_{\rm t}) = \frac{r_{\rm t}^2}{3\nu(r_{\rm t})} = \frac{M_{\rm d} (r_{\rm t})}{2\dot{M}_{\rm acc}^{\rm Tur}}.
\end{equation}

Thus, the mass loss rate becomes comparable to $ \dot{M}_{\rm acc}^{\rm Tur}$ when $t_{\rm age} \la t_{\rm vis } (r_{\rm t})$,
and no mass loss is produced from disks that have $ t_{\rm age} \ga  t_{\rm vis } (r_{\rm t})$ 
as the disks already disperse.

One may wonder what is the origin of viscosity at the outer edge of truncated disks.
While it is still poorly unconstrained,
(at least) two possibilities can be considered: MRI and hydrodynamical turbulence.
Note that MCW cannot play such a role as MHD turbulence is quenched due to non-ideal MHD effects.
If MRI operates both in the inner disk and at the outer edge, then the mass loss rate from inner winds is regulated by MRI.
If MRI or hydrodynamical turbulence operates at the outer edge, but if MCW operates in the inner disk,
then the mass loss rate from inner winds is controlled by MCW.
This is because the contribution coming from MRI at the outer edge should be negligible (see Appendix \ref{sec:app5}).
Also, the mass loss rate from outer winds is already dominated by EPE 
because $\dot{M}_{\rm loss, tru}^{\rm EPE} \simeq \dot{M}_{\rm acc}^{\rm Tur}$ and $\dot{M}_{\rm loss, tru}^{\rm MRI} < \dot{M}_{\rm acc}^{\rm Tur}$.

\section{Monte-Carlo Based, Population Synthesis Calculations} \label{sec:app5}

We conduct a parameter study to explore the effect of variation of model parameters on our results (e.g., Figure \ref{fig3}).
To proceed, we use the Monte-Carlo approach and generate disk populations.
In the approach, the values of model parameters are chosen randomly with assumed distributions.
The resulting value of $\dot{M}_{\rm loss} / \dot{M}_{\rm acc}$ is computed, using a set of these parameters.
As shown below, the variation of model parameters does not change our conclusion significantly.

\begin{figure*}
\begin{minipage}{17cm}
%\begin{figure}%[!ht]
\begin{center}
\includegraphics[height=6.8cm]{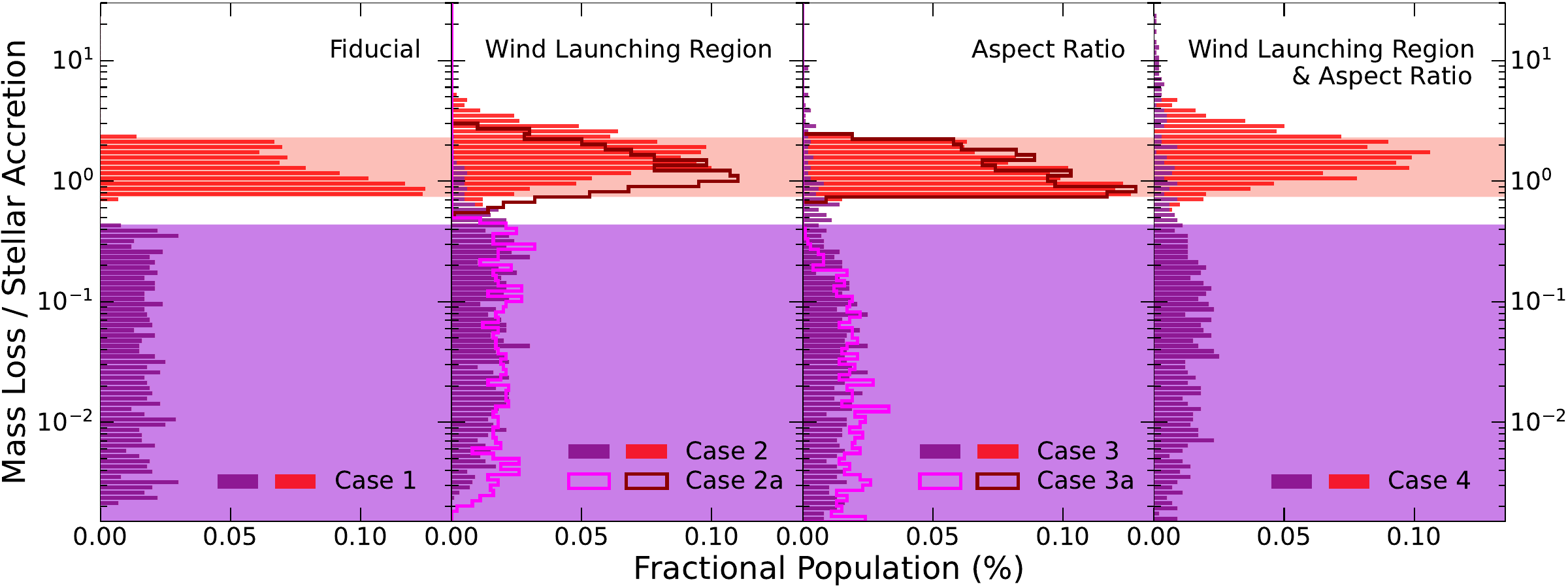}
\caption{
Disk populations for inner magnetic winds. 
The populations are generated by the Monte-Carlo approach (Table \ref{table_app5a}).
In Case 1, only the values of $r_{\rm A}/r$ and $\epsilon$ vary.
In Cases 2 and 2a, those of $r_{\rm in}$ and $r_{\rm out}$ change in addition to the values of $r_{\rm A}/r$ and $\epsilon$.
In Cases 3 and 3a, the effect of variation of $h_{0}$ is examined.
In Case 4, all the parameters are altered.
The variation of parameters leads to overlapping two populations (i.e., MCW and MRI).
However, the overlap can be disentangled, by observing an enough number of disks.
}
\label{fig5}
\end{center}
%\end{figure}
\end{minipage}
\end{figure*}

We first describe the range and distribution of model parameters adopted in this work.
For inner magnetic winds, 
four parameters ([$r_{\rm A}/r$, $r_{\rm in}$, $r_{\rm out}$, $h_0$] and [$\epsilon$, $r_{\rm in}$, $r_{\rm out}$, $h_0$]) need to be specified
both in the MCW and MRI cases, respectively (Tables \ref{table_app5a}).
Note that the dependence of $M_{\rm s}$ is explored effectively by changing $h_{0}$ in our model.
The range of the parameters is chosen, based on recent theoretical and observational studies (e.g., Figure \ref{fig1} and Sections \ref{sec:mcw} and \ref{sec:mri}).
Given that the true, underlying distribution is unknown, 
we adopt a uniform distribution in linear space for four parameters ($r_{\rm A}/r$, $r_{\rm in}$, $r_{\rm out}$, $h_0$)
and a uniform distribution in logarithmical space for one parameter ($\epsilon$);
for the latter, logarithmical space is used to equally cover a larger parameter space in Figure \ref{fig5}.
We consider that this is a conservative choice as all the plausible values of parameters are examined.

For outer thermal winds, four parameters ($M_{\rm s}$, $G_0$, $r_{\rm d}$, and $M_{\rm d}$) need to be selected (Table \ref{table_app5b}).
As with the case for inner winds, the range of parameters is picked, according to recent theoretical and observational studies (see Figure \ref{fig1} and Section \ref{sec:epe}),
and distributions are chosen to explore all the possible values of parameters.
It should be noted that the ranges are constrained by the original \textsc{fried} grid as well \citep{2018MNRAS.481..452H}.
Using these parameters, realization of disk populations is conducted 1000 times for each case.
The value of $F_{\rm FUV}$ is chosen such that disk populations distribute uniformly in logarithmic space.

We then discuss the resulting disk populations.
Figure \ref{fig5} shows the results for inner magnetic winds.
In order to elucidate the effect of each parameter, we consider six cases in total (Table \ref{table_app5a}).

Case 1 explores the variation of only $r_{\rm A}/r$, $r_{\rm in}$ and $\epsilon$.
As expected, most disks are located in the shaded regions;
since the value of $\epsilon$ is sampled uniformly in logarithmical space,
the resulting population becomes flat for the MRI case, which is beneficial for examining the effect of other parameters.

Case 2 examines how the variation of wind launching regions affects disk populations.
Our results show that both the populations spread vertically and the two cases (MCW and MRI) overlap.
This overlap originates mainly from the choice of the inner boundary of wind launching regions;
when smaller $r_{\rm in}$ is picked (see Case 2),
launching massive winds becomes possible due to a high density at the wind base for the MRI case. 
In fact, such a diffused population disappears when $r_{\rm in} =1$ au (see Case 2a).
However, the diffused population for the MRI case is not large enough, compared with the population for the MCW case.
This suggests that if an enough number of disks are observed, distinction between the MCW and MRI cases is possible.

Case 3 studies the effect of $h_0$.
We confirm a similar trend that while the populations between the MCW and MRI cases overlap due to massive winds launched from low latitudes (i.e., low $h_0$, c.f. Case 3a),
the resulting overlap can be differentiated.

In Case 4, all the parameters vary.
The corresponding population for the MRI case further extends toward higher values of $\dot{M}_{\rm loss} / \dot{M}_{\rm acc}$.
However, such a population is minor, compared with the MCW case.

In summary, it can be concluded that the shaded regions defined in Figure \ref{fig3} are reasonable for demonstration purpose in the proof-of-concept study,
and even if the variation of model parameters is taken into account, differentiation between the MCW and MRI cases is possible, by observing a good number of disks exhibiting winds.

\begin{figure*}
\begin{minipage}{17cm}
%\begin{figure}%[!ht]
\begin{center}
\includegraphics[height=4cm]{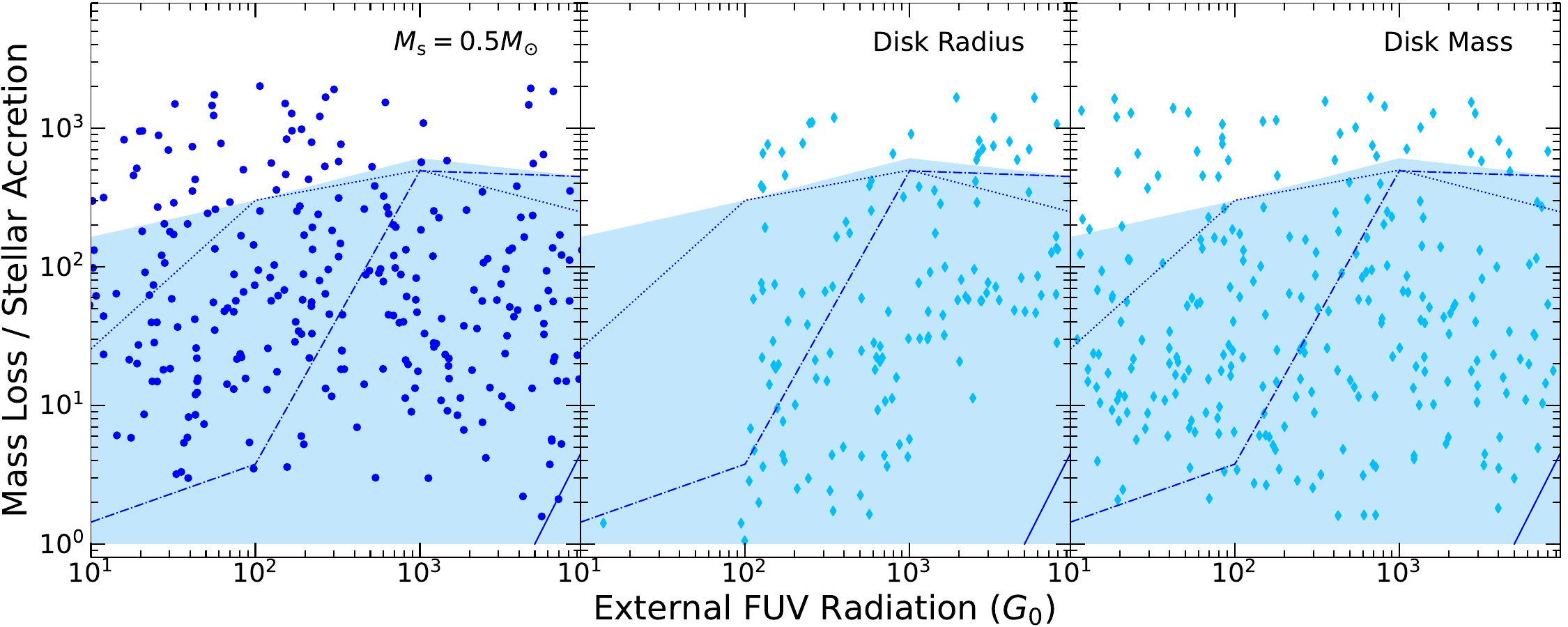}
\includegraphics[height=4cm]{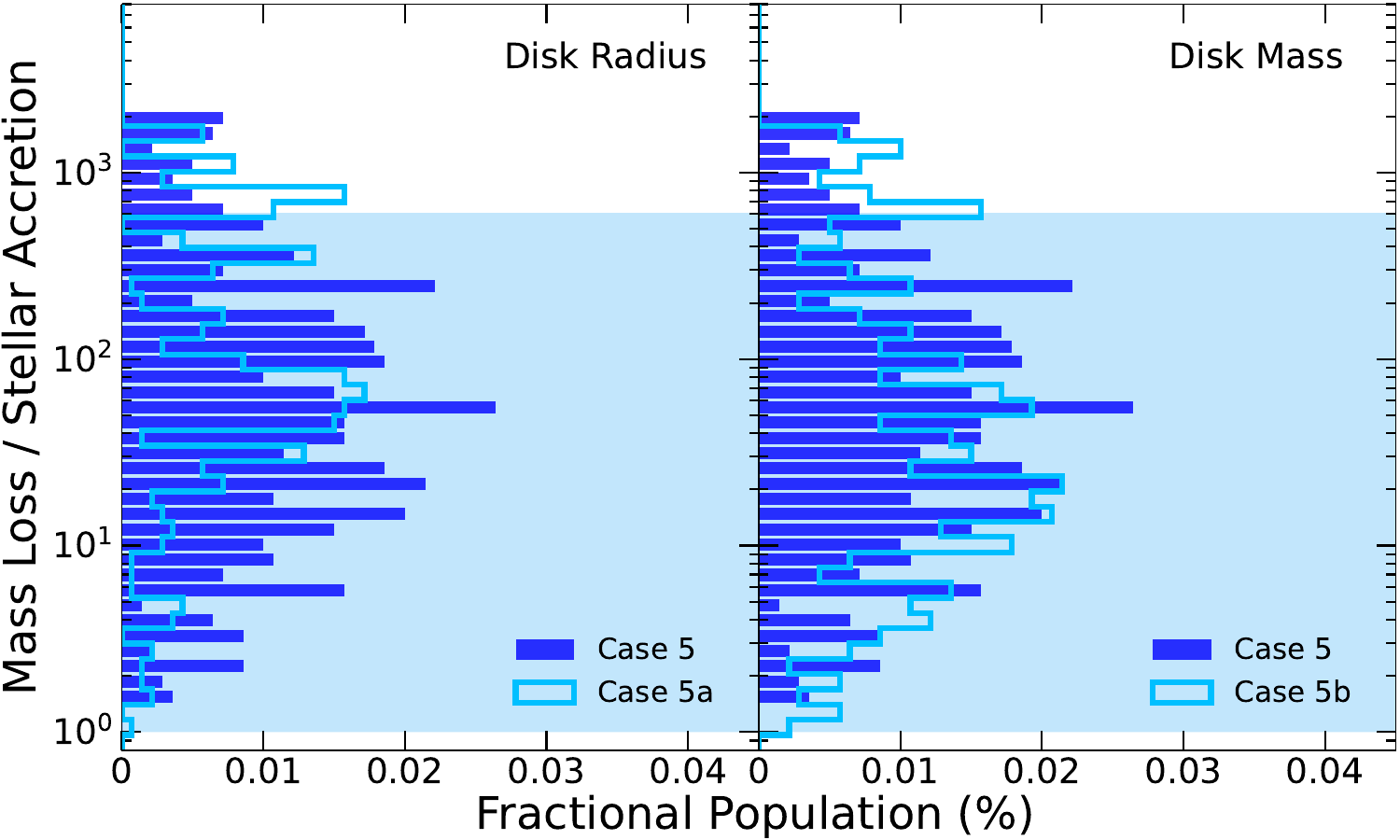}
\includegraphics[height=4cm]{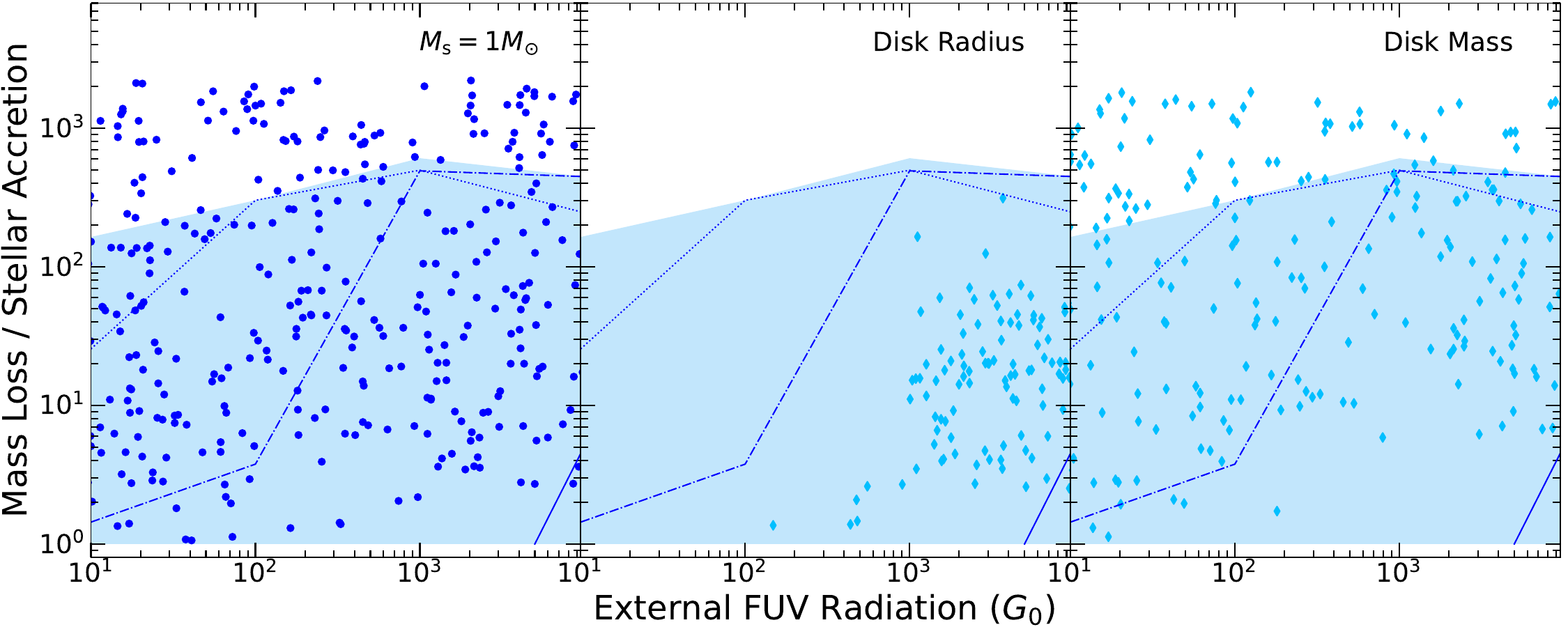}
\includegraphics[height=4cm]{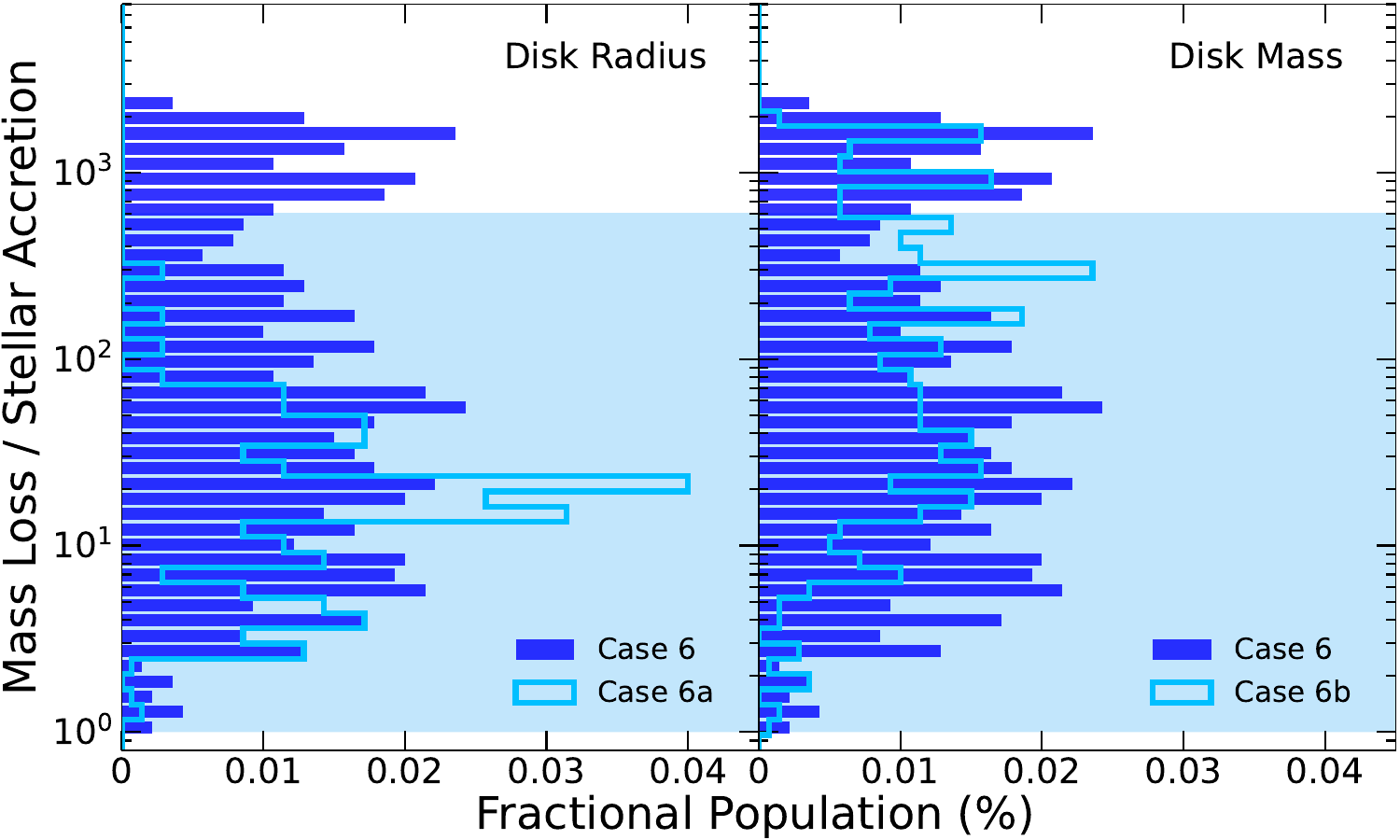}
\includegraphics[height=4cm]{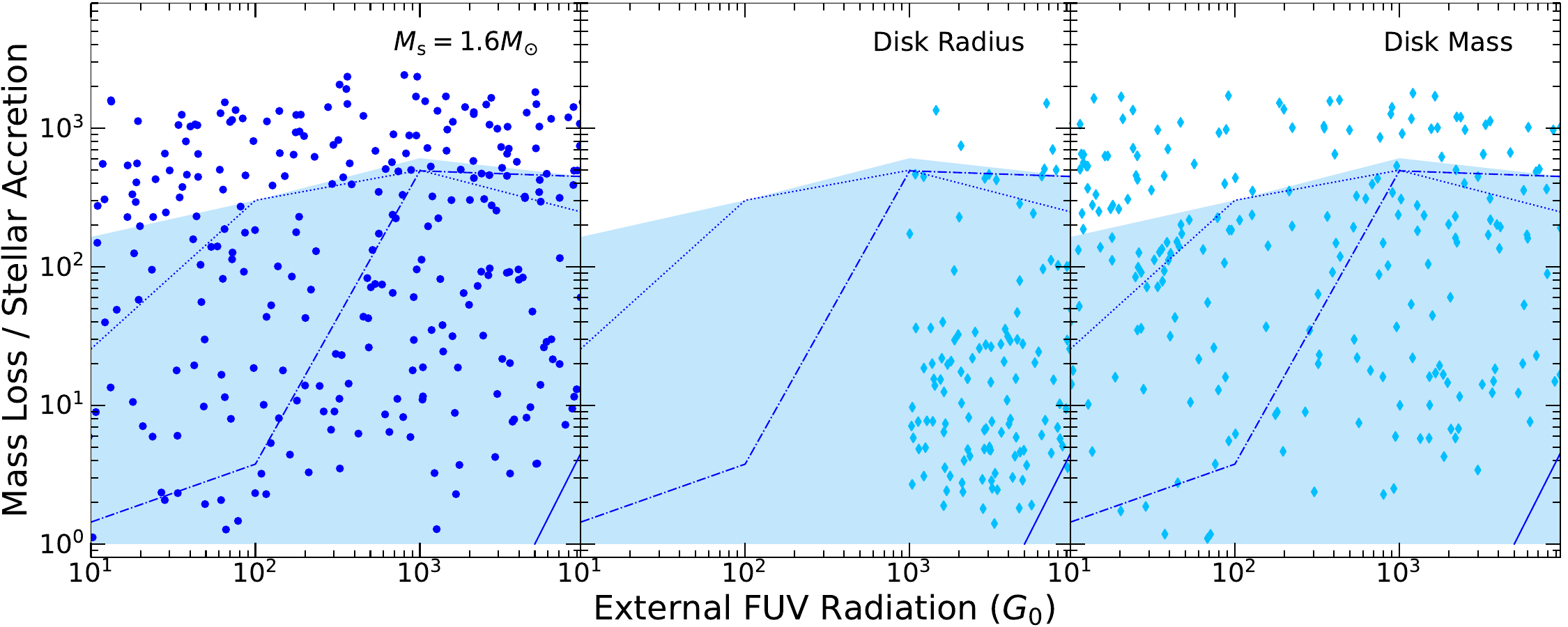}
\includegraphics[height=4cm]{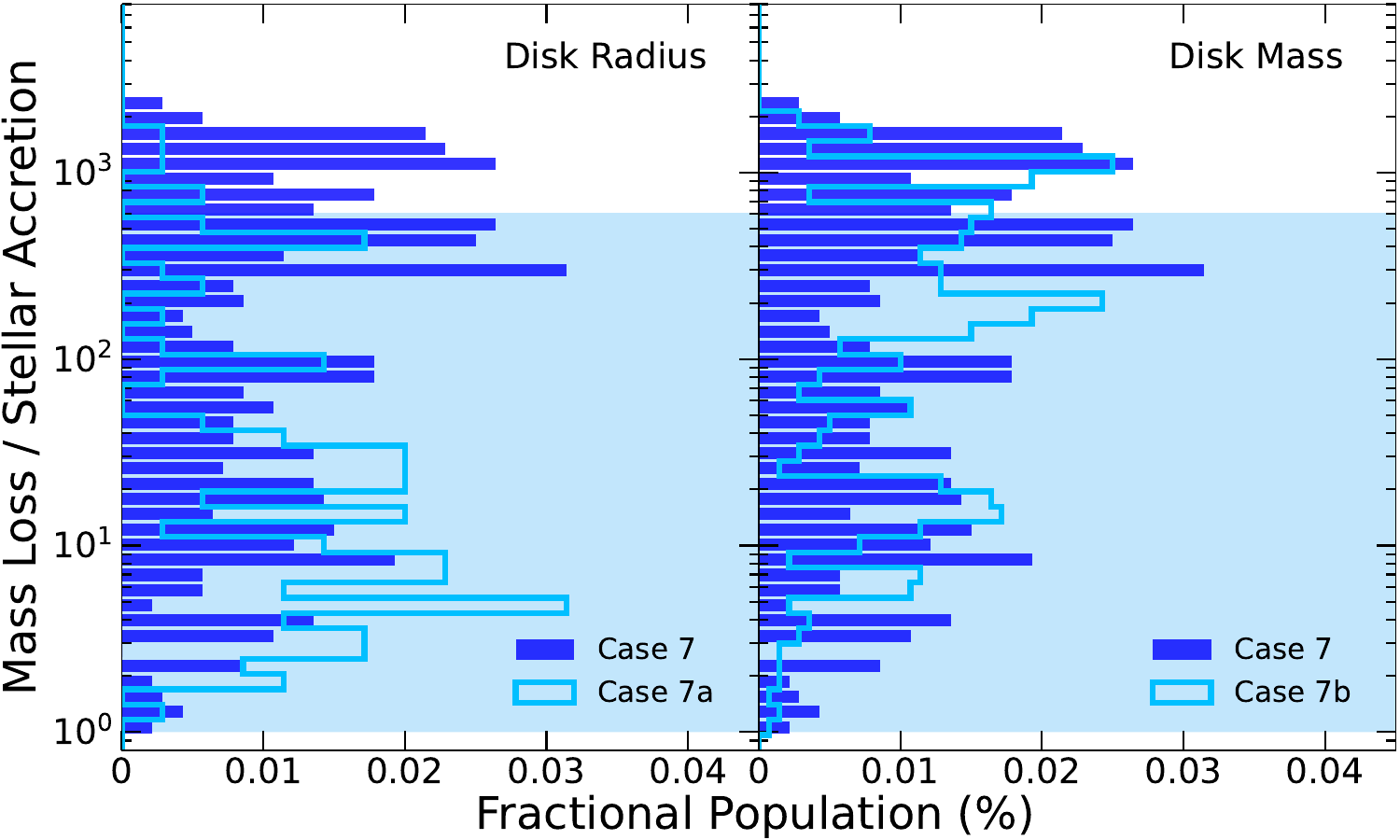}
\caption{
Disk populations for outer thermal winds. 
The populations are generated by the Monte-Carlo method (Table \ref{table_app5b}).
The top panel shows the results for the case that $M_{\rm s}=0.5 M_{\odot}$,
the middle one is for the case that $M_{\rm s}=1 M_{\odot}$,
and the bottom is for the case that $M_{\rm s}=1.6M_{\odot}$.
In the left panel, the disk populations distribute in the $\dot{M}_{\rm loss}/ \dot{M}_{\rm acc}-F_{\rm FUV}$ diagram.
Both $r_{\rm d}$ and $M_{\rm d}$ vary in all the plots; 
compared with the left plot, the ranges of $r_{\rm d}$ and $M_{\rm d}$ shrink on the central and right plots, respectively.
In the right panel, the fractional populations are shown, in order to examine the effect of the range of model parameters.
As expected, the resulting disk populations are most sensitive to the disk radius.
The blue shaded region is a good representative of disk populations for wide ranges of $r_{\rm d}$ and $M_{\rm d}$.}
\label{fig6}
\end{center}
%\end{figure}
\end{minipage}
\end{figure*}

Figure \ref{fig6} shows the results for outer thermal winds.
In our preliminary efforts, we have found that some random combinations of two parameters ($r_{\rm d}$ and $M_{\rm d}$) can lead to unrealistically high values of $\dot{M}_{\rm loss} / \dot{M}_{\rm acc}$;
it can be written as
\begin{equation}
\frac{ \dot{M}_{\rm loss} }{ \dot{M}_{\rm acc} } = \frac{ \dot{M}_{\rm loss} }{M_{\rm d}} \frac{M_{\rm d}}{ \dot{M}_{\rm acc}} \equiv \frac{\tau_{\rm acc}}{\tau_{\rm loss}},
\end{equation}
and hence such high values come from very small values of $\tau_{\rm loss}$.
Given that the disk lifetime should be determined by $\min(\tau_{\rm acc}, \tau_{\rm loss})$,
we cut off the value of $\tau_{\rm loss}$ smaller than $10^6$ yr.
In addition, we currently consider disks actively undergoing accretion.
We therefore remove disks that have $\dot{M}_{\rm acc}$ smaller than $10^{-9} M_{\odot}$ yr$^{-1}$.
In this parameter study, nine cases are examined in total (Table \ref{table_app5b}).

Cases 5, 5a, and 5b adopt that $M_{\rm s}=0.5 M_{\odot}$.
Our results show that disk populations are most sensitive to the disk size;
the nearly entire population disappears at $F_{\rm FUV} \la 10^2 G_0$ when $r_{\rm d}  \le 50$ au (Case 5a).
This occurs because winds from such tiny disks are negligible.
Disk population does not change very much even if the range of $M_{\rm d}$ varies (Case 5b).

Cases 6, 6a, and 6b consider that $M_{\rm s}=1 M_{\odot}$.
As with the case for that $M_{\rm s}=0.5 M_{\odot}$,
$r_{\rm d}$ is the most sensitive parameter;
for this case, disk population distributes at $F_{\rm FUV} \ga 10^3 G_0$ due to higher gravitational potential of the host star (Case 6a);
equivalently, higher $F_{\rm FUV}$ is needed to launch winds.

Finally, Cases 7, 7a, and 7b study that $M_{\rm s}=1.6 M_{\odot}$.
We confirm a similar trend, while the disk radius dependence is comparable to the case that $M_{\rm s}=1 M_{\odot}$.
In summary, the shaded region specified in Figure \ref{fig3} serves as a good guide to constrain disk population for the EPE case as well.

\begin{figure*}
\begin{minipage}{17cm}
%\begin{figure}%[!ht]
\begin{center}
\includegraphics[width=15cm]{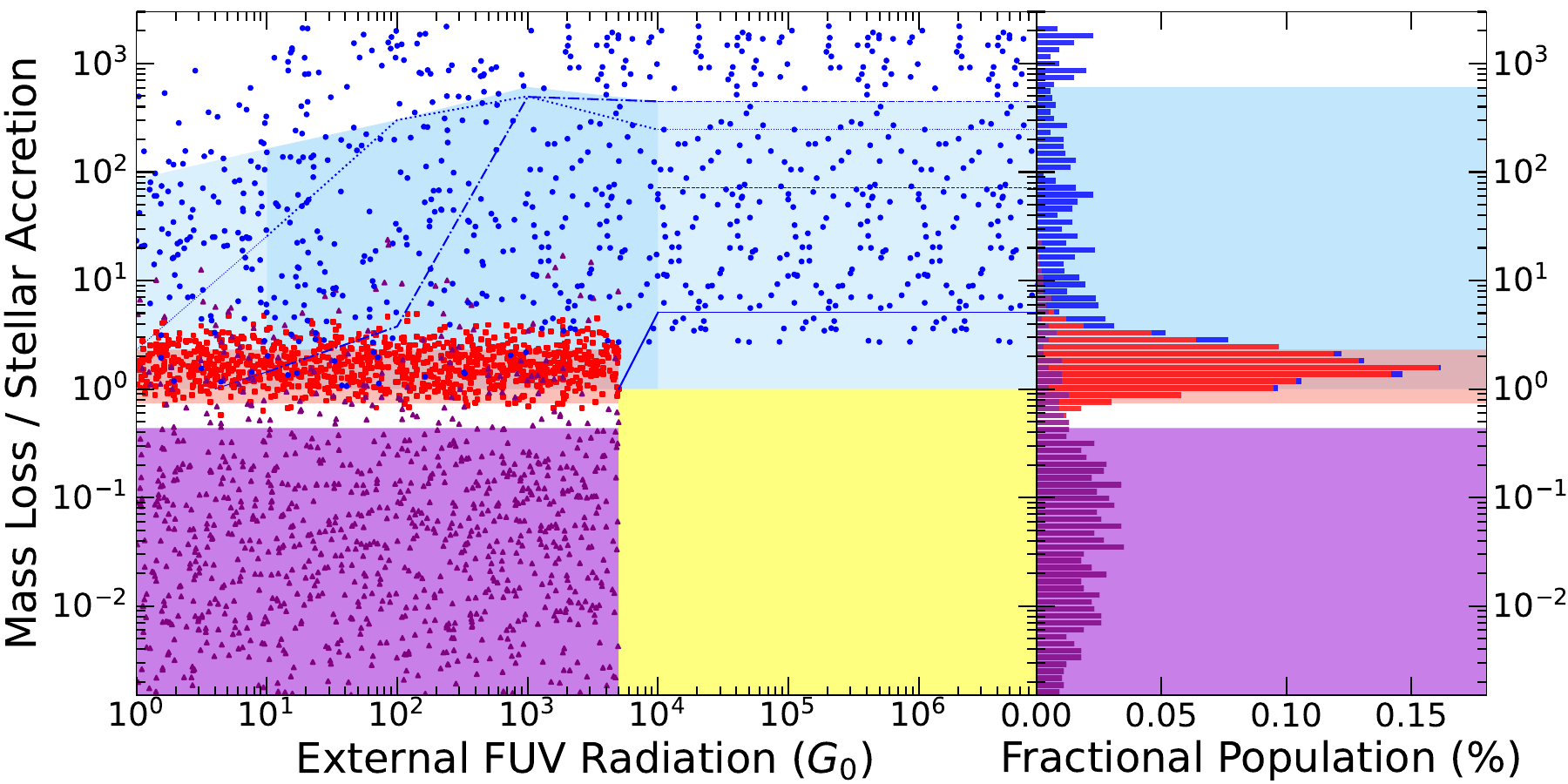}
\caption{
The synthesized results for the case that $M_{\rm s}=1 M_{\odot}$.
The shaded regions well represent the disk populations produced by the Monte-Carlo method.}
\label{fig7}
\end{center}
%\end{figure}
\end{minipage}
\end{figure*}

Figure \ref{fig7} shows the synthesized results for the case that $M_{\rm s}=1 M_{\odot}$.
It is clear that the shaded regions defined in Figure \ref{fig3} well capture disk populations generated by the Monte-Carlo methods.
Hence it is reasonable to consider that the parameter space constrained by our model serves as good reference for differentiating the origin of winds from protoplanetary disks.

%\begin{table*}
%\begin{minipage}{17cm}
\begin{sidewaystable}[h]
%\begin{table}
%\begin{center}
\centering
\caption{Sets of model parameters for inner magnetic winds}
\label{table_app5a}
{\scriptsize
\begin{tabular}{l||c|c|c|c|c|c|c|c|c|c}
\hline
                            & \multicolumn{2}{c|}{Magnetic lever arm}   & \multicolumn{2}{c|}{Proportionality constant}  & \multicolumn{2}{c|}{Inner boundary}  & \multicolumn{2}{c|}{Outer boundary}   &  \multicolumn{2}{c}{Aspect ratio}       \\            
                             & \multicolumn{2}{c|}{$r_{\rm A}/r$}            & \multicolumn{2}{c|}{$\epsilon$}               & \multicolumn{2}{c|}{$r_{\rm in}$ (au)}  & \multicolumn{2}{c|}{$r_{\rm out}$ (au)} & \multicolumn{2}{c}{$h_0$}                                                        \\ \hline \hline
                             & Range      & Distribution                          & Range & Distribution                               & Range & Distribution                         & Range & Distribution                          & Range & Distribution   \\  \hline
Case 1 (Fiducial)  & $1.2-1.6$ & Linearly uniform & $10^{-5} - 2 \times 10^{-3}$ &  Logarithmically uniform  & $1$ &                                           & $10$  &                                             & $0.05$ &    \\  \hline
Case 2                 & $1.2-1.6$ & Linearly uniform & $10^{-5} - 2 \times 10^{-3}$ &  Logarithmically uniform  & $0.05-1$ & Linearly uniform       & $5-20$  & Linearly uniform                   & $0.05$ &   \\ 
Case 2a               & $1.2-1.6$ & Linearly uniform & $10^{-5} - 2 \times 10^{-3}$ &  Logarithmically uniform  & $1$ &                                          & $5-20$  & Linearly uniform                   &  $0.05$ &   \\  \hline
Case 3                 & $1.2-1.6$ & Linearly uniform & $10^{-5} - 2 \times 10^{-3}$ &  Logarithmically uniform  &  $1$ &                                         &  $10$  &                                              & $0.01-0.1$ & Linearly uniform   \\ 
Case 3a               & $1.2-1.6$ & Linearly uniform & $10^{-5} - 2 \times 10^{-3}$ &  Logarithmically uniform  &  $1$ &                                         &  $10$  &                                              & $0.05-0.1$ & Linearly uniform   \\ \hline
Case 4                 & $1.2-1.6$ & Linearly uniform & $10^{-5} - 2 \times 10^{-3}$ &  Logarithmically uniform  & $0.05-1$ & Linearly uniform       & $5-20$  & Linearly uniform                   & $0.01-0.1$ & Linearly uniform   \\ \hline
\end{tabular}
}
%\end{center}
%\end{table}
\end{sidewaystable}
%\end{minipage}
%\end{table*}

%\begin{table*}
%\begin{minipage}{17cm}
\begin{table}
\begin{center}
\caption{Sets of model parameters for outer thermal winds}
\label{table_app5b}
{\scriptsize
\begin{tabular}{l||c|c|c|c|c|c}
\hline
                           & Stellar mass                    & External FUV radiation               & \multicolumn{2}{c|}{Disk radius}          & \multicolumn{2}{c}{Disk mass}                            \\            
                           & $M_{\rm s} (M_{\odot})$  & $F_{\rm FUV}$ ($G_0$)            & \multicolumn{2}{c|}{$r_{\rm d}$ (au)}   & \multicolumn{2}{c}{$M_{\rm d} (M_{\odot})$}    \\ \hline \hline
                           &                                         &                                                    & Range      & Distribution                        & Range & Distribution                                           \\  \hline
Case 5               & 0.5                                   & 1, 10, $10^2$, $10^3$, $10^4$   & $30-400$  & Linearly uniform                & $10^{-3} -10^{-1}$ &  Logarithmically uniform     \\  
Case 5a             & 0.5                                   & 1, 10, $10^2$, $10^3$, $10^4$   & $30-50$    & Linearly uniform                & $10^{-3} -10^{-1}$ &  Logarithmically uniform     \\  
Case 5b             & 0.5                                   & 1, 10, $10^2$, $10^3$, $10^4$   & $30-400$  & Linearly uniform                & $10^{-3} -10^{-2}$ &  Logarithmically uniform     \\  \hline
Case 6               & 1                                      & 1, 10, $10^2$, $10^3$, $10^4$   & $30-400$  & Linearly uniform                & $10^{-3} -10^{-1}$ &  Logarithmically uniform     \\  
Case 6a             & 1                                      & 1, 10, $10^2$, $10^3$, $10^4$   & $30-50$    & Linearly uniform                & $10^{-3} -10^{-1}$ &  Logarithmically uniform     \\  
Case 6b             & 1                                      & 1, 10, $10^2$, $10^3$, $10^4$   & $30-400$  & Linearly uniform                & $10^{-3} -10^{-2}$ &  Logarithmically uniform     \\  \hline
Case 7               & 1.6                                   & 1, 10, $10^2$, $10^3$, $10^4$   & $30-400$  & Linearly uniform                & $10^{-3} -10^{-1}$ &  Logarithmically uniform     \\  
Case 7a             & 1.6                                   & 1, 10, $10^2$, $10^3$, $10^4$   & $30-50$    & Linearly uniform                & $10^{-3} -10^{-1}$ &  Logarithmically uniform     \\  
Case 7b             & 1.6                                   & 1, 10, $10^2$, $10^3$, $10^4$   & $30-400$  & Linearly uniform                & $10^{-3} -10^{-2}$ &  Logarithmically uniform     \\ 
\hline
\end{tabular}
}
\end{center}
\end{table}
%\end{minipage}
%\end{table*}

\section{Estimating Inner wind Mass Loss Rates for observed systems from [OI] 6300 {\AA}} \label{sec:app3}

We here consider observed systems and estimate the mass loss rate that very likely originates from inner magnetic winds, using the forbidden [OI] 6300{\AA} emission line.
Previous observations find strong evidence that the narrow/low velocity component (LVC) of the [OI] 6300{\AA} emission 
traces the warm/hot neutral (atomic) layer of a slow, molecular/neutral disk wind launched from the inner part of protoplanetary disks 
\citep[e.g.,][]{2014A&A...569A...5N,2016ApJ...831..169S,2021ApJ...913...43W}.
We therefore focus on the component in the following analysis.

In principle, the wind mass loss rate ($\dot{M}_{\rm loss}^{\rm in}$) is estimated as
\begin{equation}
    \dot{M}_{\rm loss}^{\rm in} \sim M_{\rm wind} \frac{v_{\rm wind}}{L_{\rm wind}},
\end{equation}
where $M_{\rm wind}$ is the total mass carried away by winds, $v_{\rm wind}$ the wind velocity, and $L_{\rm wind}$ the physical extent occupied by the winds. 
Observations of the [OI] 6300{\AA} emission line provide its line flux and shape.
Thus, one can estimate $\dot{M}_{\rm loss}^{\rm in}$ by converting these two observables to three physical quantities.

The quantity, $v_{\rm wind}$, is a direct measurement of the velocity shift of the [OI] 6300{\AA} narrow line component with respect to the system radial velocity,
The quantities, $M_{\rm wind}$ and $L_{\rm wind}$, are proportional to the total dereddened [OI] 6300{\AA} narrow component line luminosity (L[OI]).
We follow a simplified method outlined in \citet{2014A&A...569A...5N} to derive $M_{\rm wind}$ from L[OI];
a more elaborate method can be found elsewhere \citep[e.g.,][]{2018ApJ...868...28F}.
More specifically, we use equation (5) in \citet{2014A&A...569A...5N},
assuming that $j([{\rm OI}] 6300)$ = 6.0$\times$10$^{-16}$ $\pm$ 5.0$\times$10$^{-16}$ erg s$^{-1}$ OI$^{-1}$ and $\alpha({\rm OI})$ = 5.0$\times$10$^{-4}$ $\pm$ 3.0$\times$10$^{-5}$, 
where $j([{\rm OI}] 6300)$ is the line emissivity and $\alpha({\rm OI})$ is the fraction of the neutral gas wind made up of OI. 
Estimating $L_{\rm wind}$ is most unconstrained because it depends sensitively on the geometry of winds.
We here consider two different geometries, which may provide upper and lower bounds to the possible (realistic) ranges of $L_{\rm wind}$. 
The first geometry of winds is spherical symmetric.
This is adopted in \citet{2014A&A...569A...5N} and leads to the most conservative estimate for  $L_{\rm wind}$,
which gives a lower limit .
The second geometry is a bipolar cone.
By assuming that winds are launched from the innermost ($\sim$ 0.1 au) region of the disk,
the resulting estimate of $L_{\rm wind}$ gives an upper limit.
For both cases of geometry, the volumetric density of [OI] is calculated, using equation (7) in \citet{2014A&A...569A...5N}, 
and then this density is converted to $L_{\rm wind}$, using the appropriate geometric conversion factor for these cases. 
We use the mid-point between the two geometries for $L_{\rm wind}$ in our estimates, 
with the end points (upper and lower limits from the conical and spherical geometries, respectively) constituting the 1-$\sigma$ errors on $L_{\rm wind}$. 

All errors in variables required to compute $\dot{M}_{\rm loss}^{\rm in}$ are assumed to be uncorrelated and therefore follow traditional error propagation rules. 
To compute the ratio of $\dot{M}_{\rm loss}^{\rm in}$ to $\dot{M}_{\rm acc}$ and its error, we assume an order of magnitude error range on the measured accretion rates. 
Since it is very unlikely that simultaneous observations were conducted to determine both $\dot{M}_{\rm loss}^{\rm in}$ and $\dot{M}_{\rm acc}$,
it would be safest to apply this error range.
We also assume that the measurements of $\dot{M}_{\rm loss}^{\rm in}$ and $\dot{M}_{\rm acc}$ are intrinsically correlated in nature,
that is, it is expected that a higher $\dot{M}_{\rm acc}$ leads to a higher value of $\dot{M}_{\rm loss}^{\rm in}$.
Therefore, we assume that errors associated with $\dot{M}_{\rm loss}^{\rm in}$ and $\dot{M}_{\rm acc}$ are correlated, and adopt error propagation for correlated variables. 

Our estimates for observed systems are summarized in Table \ref{table_app3};
observed data needed to compute $\dot{M}_{\rm loss}^{\rm in}/\dot{M}_{\rm acc}$ and its error are adapted from \citet{2016ApJ...831..169S,2018A&A...609A..87N,2020A&A...643A..32G}.
The value of $F_{\rm FUV}$ is estimated as done in Section \ref{sec:app4}.

%\begin{table*}
%\begin{minipage}{17cm}
\begin{table}
\begin{center}
\caption{The ratio of the inner wind mass loss rate to the stellar accretion rate for observed systems}
\label{table_app3}
%{\scriptsize
\begin{tabular}{lcc}
\hline
Target Name   &  $F_{\rm FUV} (G_0)$  & $\dot{M}_{\rm loss}^{\rm in} / \dot{M}_{\rm acc}$         \\   \hline      
CI Tau             & 26                                  & 0.016 $\pm$ 0.38 \\
CW Tau           & 40                                  & 0.018 $\pm$ 0.61 \\
DG Tau           & 25                                  & 0.099 $\pm$ 1.8 \\
DL Tau            & 32                                  & 0.032 $\pm$ 0.75 \\
DN Tau           & 4                                    & 0.26 $\pm$ 4.9 \\
DR Tau           & 29                                  & 0.012 $\pm$ 0.31 \\
GI Tau            & 41                                  & 0.063$\pm$ 2.0 \\
HN Tau           & 38                                 & 0.40 $\pm$ 7.4 \\
HQ Tau           & 37                                 & 0.33 $\pm$ 6.2 \\
Ex Lup            & 5                                   & 0.0041 $\pm$0.085 \\
RU Lup           & 5                                  & 0.01808 $\pm$ 0.35 \\
RY Lup           & 9                                   & 0.022 $\pm$ 0.54 \\
Sz 73              & 9                                   & 0.29 $\pm$ 5.5 \\
Sz 98              & 3                                   & 0.012 $\pm$ 0.27 \\
Ass Cha T 2-3 & 2                                 & 0.18 $\pm$ 4.0 \\
TW Cha          & 6                                  & 0.17 $\pm$3.9 \\
CT Cha A        & 3                                 & 0.015 $\pm$ 0.37 \\
Sz 22              & 6                                & 0.0163 $\pm$ 0.38 \\
VW Cha          & 3                                & 0.11 $\pm$ 2.5 \\
ESO H$\alpha$ 562 & 5                       & 1.6 $\pm$ 35 \\
Ass Cha T 2-38 & 5                              & 0.15 $\pm$ 3.3 \\
CHXR 79           & 7                               & 5.5 $\pm$ 123 \\
Ass Cha T 2-40 & 2                                & 0.00021 $\pm$ 0.11 \\
Sz 32                 & 7                              & 0.024 $\pm$ 0.56 \\
Ass Cha T 2-44 & 1                              & 0.0023 $\pm$ 0.054 \\
Sz 37                  & 7                             & 0.042 $\pm$ 0.99 \\
CHX 18N            & 7                             & 0.099 $\pm$ 2.3 \\
Ass Cha T 2-52   & 6                            & 0.0074 $\pm$  0.17 \\
CW Cha              & 1                            & 0.011 $\pm$ 0.27 \\
CoKu Tau 4         & 48                          & 7.9 $\pm$ 166 \\
FZ Tau                & 61                           & 0.015 $\pm$ 0.37 \\
 \hline 
\end{tabular}
%}
\end{center}
\end{table}
%\end{minipage}
%\end{table*}

\section{Estimating Outer wind Mass Loss Rates for observed systems} \label{sec:app4}

The outer wind mass loss rate ($\dot{M}_{\rm loss}^{\rm out}$) for observed systems can be estimated as \citep[e.g.,][]{1998ApJ...499..758J}
\begin{equation}
 \label{eq:Mloss_out}
            \left( \frac{ \dot{M}_{\rm loss}^{\rm out} }{ 10^{-8} M_{\odot} \mbox{ yr}^{-1}} \right) 
\simeq \left( \frac{R_{\rm LF}}{1200 \mbox{ au}} \right)^{3/2} \left( \frac{d}{1 \mbox{ pc}} \right)^{-1} \left( \frac{\dot{N}_{\rm Ly}}{10^{45} \mbox{ s}^{-1}} \right)^{1/2}, 
\end{equation}
where $R_{\rm LF}$ is the radius of the ionization front defined around the surface of a protoplanetary disk by the external ionizing source,
$d$ is the distance between the disk and the source, and $\dot{N}_{\rm Ly}$ is the extreme UV (EUV) photon count from the source.
As done in Section \ref{sec:epe}, the disk accretion rate ($\dot{M}_{\rm acc}$) onto the host star is estimated from the disk mass.
Thus, the ratio of $\dot{M}_{\rm loss}^{\rm out}$ to $\dot{M}_{\rm acc}$ for observed systems is estimated from equations (\ref{eq:Mloss_out}) and (\ref{eq:corr_Macc_Md}).

We also estimate the external FUV field ($F_{\rm FUV}$) as
\begin{equation}
F_{\rm FUV} = F_{\rm FUV}(\mbox{at 0.1 pc}) \left( \frac{d}{0.1 \mbox{ pc}} \right)^{-2}.
\end{equation}

Our estimates for observed systems are summarized in Table \ref{table_app4}.
We obtain the observed data from \citet{1998AJ....116..322H,2018ApJ...860...77E,2021MNRAS.501.3502H}
and conduct traditional error propagation calculations to estimate errors.
When errors of the data are not provided,
we assume that errors of $R_{\rm LF}$ are $\pm$ 10 \% of the observed value,
errors of $d$ are $\pm$ 50 \% of the observed value,
and errors of $\dot{N}_{\rm Ly}$ are $\pm$ 2 times the observed value.

%\begin{table*}
%\begin{minipage}{17cm}
\begin{table}
\begin{center}
\caption{The ratio of the outer wind mass loss rate to the stellar accretion rate for observed systems}
\label{table_app4}
%{\scriptsize
\begin{tabular}{lcc}
\hline
Target Name                             &  $F_{\rm FUV} (10^5 G_0)$  & $\dot{M}_{\rm loss}^{\rm out} / \dot{M}_{\rm acc}$         \\   \hline      
NGC 2024 Proplyd 1 (VLA 1)    & 2.4	                                         & 5.1 $\pm$ 6.7 \\
NGC 2024 Proplyd 2 (VLA 4)	& 5.4	                                         & 52 $\pm$ 63    \\
NGC 2024 Proplyd 3 (VL A 20b) & 5.5                                       & 35 $\pm $ 48     \\
NGC 2024 Proplyd 4 (VL A 20a)  &6.2                                       & 19 $\pm$ 24        \\
NGC 2024 Candidate Proplyd 5  & 1.2                                      &15 $\pm$	20      \\
NGC 2024 Candidate Proplyd 7  &1.8                                       & 9.3 $\pm$ 11      \\
ONC 152-319                              & 3.1                                      & $1.8 \times 10^2 \pm 2.7 \times 10^2$      \\
ONC 154-324	                           & 8.2	                                & $1.3 \times 10^2 \pm 5.6 \times 10^2$     \\
ONC 155-338	                           & 2.8	                                & 27 $\pm $ 35     \\
ONC 159-338	                           & 15	                                & 41 $\pm $ 59	  \\
ONC 159-350	                           & 5.4	                                & 15 $\pm$	16       \\
ONC 161-314	                           &1.3                                        & 33 $\pm$ 63      \\
ONC 161-328	                           & 40   	                                & $2.2 \times 10^2 \pm 2.9 \times 10^2$   \\
ONC 166-316	                           & 24	                                &$3.1 \times 10^2 \pm 2.4 \times 10^3$    \\
ONC 168-328	                           & 27	                                & 40 $\pm$ 76     \\
ONC 170-337	                          &13	                                        & 4.2 $\pm$ 17     \\
ONC 171-340	                          & 6.4 	                               & 38 $\pm 1.2 \times 10^2$	 \\
ONC 173-341	                          & 2.3 	                               & 23 $\pm$	53     \\
ONC 177-341	                          & 5.3		                      & $1.8 \times 10^2 \pm 4.9 \times 10^2$   \\
ONC 180-331	                          & 5.4	                              & $1.5 \times 10^2 \pm 3.3 \times 10^2$   \\
 \hline 
\end{tabular}
%}
\end{center}
\end{table}
%\end{minipage}
%\end{table*}

\end{document}